\documentclass[10pt,journal,compsoc]{IEEEtran}
\IEEEoverridecommandlockouts
\usepackage{booktabs}
\usepackage{hhline}

\usepackage{cite}
\usepackage{amsmath,amssymb,amsfonts}
\usepackage{algorithmic}
\usepackage{inconsolata}
\usepackage{graphicx}
\usepackage{textcomp}
\usepackage{xcolor}
\usepackage{subfigure}
\usepackage{amsthm}
\usepackage{url}
\usepackage{threeparttable}
\usepackage{framed}
\usepackage{multicol}
\usepackage{mdwlist}

\definecolor{BrickRed}{RGB}{178,34,34}

\makeatletter  
\newif\if@restonecol  
\makeatother

\usepackage[
  n,
  operators,
  advantage,
  sets,
  adversary,
  landau,
  probability,
  notions,  
  ff,
  mm,
  primitives,
  events,
  complexity,
  asymptotics,
  keys]{cryptocode}

\makeatletter
\def\ps@IEEEtitlepagestyle{%
  \def\@oddfoot{\mycopyrightnotice}%
  \def\@evenfoot{}%
}
\def\mycopyrightnotice{%
  {\footnotesize 
  \begin{minipage}{\textwidth}
  \centering
  \textbf{This work has been submitted to the IEEE for possible publication. Copyright may be transferred without notice, after which this version may no longer be accessible.}
  \end{minipage}
    \hfill}% <--- Change here
  \gdef\mycopyrightnotice{}% just in case
}

\begin{document}
% paper title
\title{An Adaptive and Modular Blockchain Enabled Architecture for a Decentralized Metaverse}
%\title{A Decentralized Metaverse Architecture Enabled by an Adaptive and Modular Blockchain}
% \title{A Decentralized Metaverse on Adaptive and Modular Blockchain}

	\author{Ye Cheng, Yihao Guo, Minghui~Xu,~\IEEEmembership{Member,~IEEE,}
    Qin~Hu,~\IEEEmembership{Member,~IEEE,} 
    Dongxiao Yu,~\IEEEmembership{Senior Member,~IEEE,} Xiuzhen~Cheng,~\IEEEmembership{Fellow,~IEEE}
		\thanks{Corresponding author: Minghui Xu.}
		\thanks{Y. Cheng, Y. Guo, M. Xu, D. Yu, and X. Cheng are with the School of Computer Science and Technology, Shandong University, Qingdao, Shandong, China (e-mail: \{yech@mail., yhguo@mail., mhxu@, dxyu@, xzcheng@\}sdu.edu.cn).}
		\thanks{Q. Hu is with the Department of Computer and Information Science, Indiana University-Purdue University Indianapolis, USA (e-mail: qinhu@iu.edu).}
  
	}
\IEEEtitleabstractindextext{%
\begin{abstract}
A metaverse breaks the boundaries of time and space between people, realizing a more realistic virtual experience, improving work efficiency, and creating a new business model. Blockchain, as one of the key supporting technologies for a metaverse design, provides a trusted interactive environment. However, the rich and varied scenes of a metaverse have led to excessive consumption of on-chain resources, raising the threshold for ordinary users to join, thereby losing the human-centered design. Therefore, we propose an adaptive and modular blockchain-enabled architecture for a decentralized metaverse to address these issues. The solution includes an adaptive consensus/ledger protocol based on a modular blockchain, which can effectively adapt to the ever-changing scenarios of the metaverse, reduce resource consumption, and provide a secure and reliable interactive environment. In addition, we propose the concept of Non-Fungible Resource (NFR) to virtualize idle resources. Users can establish a temporary trusted environment and rent others' NFR to meet their computing needs. Finally, we simulate and test our solution based on XuperChain, and the experimental results prove the feasibility of our design.

\end{abstract}

% Note that keywords are not normally used for peerreview papers.
}

% make the title area
\maketitle

\IEEEdisplaynontitleabstractindextext
\IEEEpeerreviewmaketitle

%TODO-----------------Yihao
\section{Introduction} \label{sec:introduction}

%元宇宙的现状
The notion of metaverse was first introduced in the 1992 science fiction novel ``Snow Crash'' by Neal Stephenson. In recent years, this novel concept has gained increasing attention in various fields, from industry to academia~\cite{ning2021survey}. Researchers and developers have been exploring the potential technologies to recreate the immersive and interactive virtual world described in the novel. The goal is to realize a seamless integration of the virtual and the physical worlds, enabling new forms of communication, collaboration, and creativity~\cite{wang2022survey}. As such, the development of metaverse has significant implications for various industries such as gaming and e-commerce, as well as for social and cultural practices~\cite{lee2021all, dionisio20133d}. 

%区块链扮演了重要作用
Blockchain technology~\cite{nakamoto2008bitcoin} has achieved great success in the field of cryptocurrency due to its decentralization, transparency, and immutability, and it has been widely regarded as one of the essential technologies to realize a metaverse~\cite{wang2022survey}. 
Many studies~\cite{fu2022survey, xu2022trustless} have investigated the role of blockchain  within a metaverse and recognized that blockchain-enabled metaverse can provide a trusted environment for users who do not trust each other in the metaverse.  
However, besides the aforementioned advantages, a blockchain-enabled metaverse still has two key challenges that need to be addressed. 

%区块链共识-问题一
The first challenge is the incompatibility between the dynamic nature of the metaverse and the static consensus/ledger mechanism inherent to a current blockchain system. Specifically, as the scenes and participants within a metaverse undergo constant changes, the security and system performance requirements for the metaverse also vary accordingly. 
For example, users may exhibit higher levels of trust when the scenes are limited to a single company, but their trust diminishes when scenes require collaboration among multiple companies. In comparison, 
%the consensus mechanism and ledger in a blockchain are fixed. When a blockchain is initialized, it comes equipped with predefined consensus algorithms and ledger structures, which is suitable for the design and development of a blockchain but not ideal for dynamically changing scenarios like the metaverse.
the consensus mechanism and ledger in a blockchain are typically static. During initialization of the blockchain, specific consensus algorithm and ledger structure are predetermined, which are suitable for the initial development. However, they are not well-suited for dynamically changing scenarios in the metaverse. 
XuperChain~\cite{XuperChain} and Ethereum~\cite{wood2014ethereum} have attempted to dynamically replace the consensus algorithm of a blockchain, but they are dependent on human interventions and have issues such as high delays and low fault-tolerance. 

The \emph{Consensus/Ledger Adaptation (CLA) problem} arises due to the complex coupling design of a blockchain system, making it difficult to dynamically change its consensus algorithm and ledger structure. Human-driven strategies such as those taken by XuperChain and Ethereum face difficulties in terms of time consumption as they need to keep up with the fast-paced changes within the metaverse. 
Moreover, relying on active human participation introduces artificial behaviors, posing threats to the security of the system. To address this challenge, an adaptive consensus/ledger replacement scheme needs to be designed to eliminate the effect of human factors and cater to the changing demands of the metaverse.

%资源门槛太高-问题二
Second, the increasing demand on computing power in a metaverse has raised an entry barrier for users. Numerous metaverse scenes, including gaming~\cite{Roblox} and industrial manufacturing~\cite{Omniverse}, require significant computational power to provide immersive experiences and precise models. However, users who lack adequate computational resources face difficulties in participating, hindering the widespread adoption of the metaverse. At present, 
users can acquire computational power through hardware upgrades or by renting servers. However, the former approach involves significant asset investment, and the latter benefits the giant centralized service providers while limits the opportunity for the small ubiquitous computational power providers. %, resulting in a significant accumulation of idle resources. 
One reason resulting in such a limitation is due to the absence of a trusted distributed trading platform. 
%users can obtain computational power through hardware upgrades or server rental/sales. The former incurs high asset requirements, while the latter enriches centralized service providers and stifles individual computational power providers due to the lack of a trust distributed trading platform. 
This problem is known as the \emph{Resource Consolidation (RC) problem}, whose solution can help to utilize idle or under-utilized resources hooked on the Internet. To overcome it, a trusted distributed transaction environment based on blockchain is necessary; nevertheless, blockchain cannot directly trade physical computational resources. Therefore, a resource virtualization solution needs to be developed that can fulfill the on-chain transaction requirements and meet the computational power demands of metaverse participants.  
%enriches centralized service providers, exploiting user trust, and 
%我们的做法：
%\textcolor{blue}{Ye-TODO:check contributions    ---Yihao 0312}

%总结贡献
In this paper, we present an adaptive and modular blockchain-enabled architecture for a decentralized metaverse, aiming to effectively tackle the above two challenges. More specifically, to address the CLA problem, we propose the dynamic and adaptive consensus/ledger replacement protocols leveraging a modular blockchain to  minimize the costs associated with consensus and ledger replacement. Additionally, we incorporate AI models to ensure adaptability to the constantly changing metaverse environment with human intervention.
To solve the RC problem, we introduce the concept of Non-Fungible Resource (NFR) and propose a protocol for NFR virtualization. This protocol allows users to upload their idle resources onto the blockchain for trading. Furthermore, we implement a temporary trusted environment to reduce on-chain resource consumption. 

Our contributions are highlighted as follows: 
\begin{itemize}
\item We introduce an adaptive and modular blockchain-enabled architecture for a decentralized metaverse, highlighting the crucial role of blockchain in its development.

\item  To solve the CLA problem, we propose adaptive protocols for consensus and ledger replacements. These protocols enable the dynamic adaptation to the constantly-changing metaverse environment.  

\item To solve the RC problem, we present the concept of NFR via which computing resources can be leased in the metaverse, successfully realizing the resource virtualization and effectively providing support to the nodes in need of computational resources. 

\item We build a prototype of the mertaverse architecture to demonstrate our design and validate its performance, which can contribute to the metaverse community and help developers in designing practical metaverse applications. 	
\end{itemize}

%文章组织。  
%\textcolor{blue}{Yihao-TODO:add content    ---Yihao 0312}

%
The rest of the paper is organized as follows. 
Section~\ref{sec:related work} provides a comprehensive review on the most relevant works. 
In Section~\ref{sec:Preliminaries}, we present the preliminary knowledge necessary to understand the subsequent sections.
Section~\ref{sec:Metaverse} delves into details of the proposed blockchain-enabled metaverse architecture, whose prototype implementation and performance evaluation are discussed in Section~\ref{sec:Implementation}.

%We first review the most related works in Section~\ref{sec:related work}. 
%Then, we introduce the preliminary knowledge in Section~\ref{sec:Preliminaries}.
%In Section~\ref{sec:Metaverse}, we detail the design of adaptive decentralized metaverse. 
%Section~\ref{sec:Implementation} reports the implementation and performance evaluation. Finally, we provide concluding remarks in Section~\ref{sec:conclusion}.  

\section{Related Works and Motivations} \label{sec:related work}
%Metaverse is in an early stage of exploration. 
In this section, we first summarize the status quo of metaverse in both academia and industry, then conduct an investigation on the roles that blockchain can play in a metaverse, and finally describe the motivations behind our design on a decentralized metaverse. 

\subsection{Metaverse in Industry and Academia} 
Efforts to develop metaverse technologies in industry primarily focus on exploring its feasibility in various fields. In an online office, a metaverse allows users to display 3D physical models in a virtual space and provide strong interaction capabilities, thereby improving user engagement and work efficiency. For examples, Microsoft Mesh~\cite{MicrosoftMesh} and Horizon Workroom~\cite{HorizonWorkroom} enable multi-person meetings, while NVIDIA Omniverse~\cite{Omniverse} allows real-time collaborative 3D modeling. 
In the gaming industry, a metaverse offers users an authentic gaming experience, rich game content and diverse gameplay. Users no longer are just participants in a game but also become game designers and developers. Examples include Second Life~\cite{secondlife}, Minecraft~\cite{Minecraft}, and Roblox~\cite{Roblox}, which aim to strike a balance between game authenticity and user openness. 

In academia, researchers focus on different aspects of a metaverse such as architecture and security. They also conduct experiments to evaluate scenario usability and propose solutions to address various challenges.  
Radoff~\cite{Jon2021metaverse} introduced a seven-layer metaverse architecture.
Nair~\textit{et al.}~\cite{nair2022going} proposed a privacy-preserving framework to enhance the security of VR devices. 
Duan~\textit{et al.}~\cite{duan2021metaverse} presented a metaverse prototype for university campuses. %In CUHKSZ~\cite{duan2021metaverse}, a prototype for university campuses was presented. 
Wang~\textit{et al.}~\cite{wang2022survey} reviewed the security and privacy issues of major metaverse solutions. 
Kye~\textit{et al.}~\cite{kye2021educational} investigated the opportunities and challenges of a metaverse in education.

% In academia, some works focus on investigating and prospecting the role of specific technologies in the metaverse. Kye~\textit{et al.}~\cite{kye2021educational} investigated the opportunities and challenges of metaverse in education. Wang~\textit{et al.}~\cite{wang2022survey} focused on the security and privacy issues of major metaverse solutions. Xu~\textit{et al.}~\cite{xu2022full} focused on edge-enabled metaverse. 
% In addition, some researchers are developing new metaverse architectures and optimizing the metaverse in specific ways. For example, Jon Radoff~\cite{Jon2021metaverse} introduced a seven-layer metaverse architecture, while CUHKSZ~\cite{duan2021metaverse} implemented a university campus prototype using a consortium blockchain. Nair~\textit{et al.}~\cite{nair2022going} proposed an privacy-perserving framework to enhance the security and privacy of VR devices. 

\subsection{Blockchain for Metaverse} \label{sub-sec: bc-metaverse}
%Blockchain technology is crucial in providing trust and has supported many platforms and academic research related to the metaverse. 
Blockchain can provide a decentralized and trusted interactive environment for a metaverse. 
Decentraland~\cite{Decentraland} is a decentralized platform based on Ethereum, 
enabling users to obtain rewards through NFT creation, trade, and use/consumption. 
Cryptovoxels~\cite{Cryptovoxels} provides a highly open virtual world on Ethereum, where players can unleash their creative potential and craft their own custom items. 
Yang~\textit{et al.}~\cite{yang2022fusing} emphasized the importance of AI and blockchain in a metaverse. 
Huynh~\textit{et al.}~\cite{huynh2023blockchain} studied the role of blockchain in a metaverse from a technical perspective. 
Fu~\textit{et al.}~\cite{fu2022survey} conducted a comprehensive review on the use of blockchain and intelligent networking in creating immersive metaverse experiences.  
Huang~\textit{et al.}~\cite{huang2022fusion} discussed the potential of combining metaverse and blockchain via the  building information modeling technology
to transform the physical world into an exciting digital one.  
Metarepo~\cite{ersoy2023blockchain} leverages blockchain technologies to provide a secure mechanism for storing the digital assets in a metaverse.  
By integrating artificial intelligence and blockchain, 
the metaverse proposed by Ali~\textit{et al.}~\cite{ali2023metaverse} can provide secure and efficient healthcare services. 
Xu~\textit{et al.}~\cite{xu2022trustless} presented a blockchain-based metaverse framework to integrate hardware and software resources, stressing blockchain as the underlying layer that provides a trusted environment with high security and privacy. 

\subsection{Motivations}
Based on the above summary and analysis, one can see that the development of metaverse in industry is still in its early stage and is currently limited to specific application scenarios, such as online collaboration and gaming. 
In academia, research efforts are largely dedicated to the proposals on innovative concepts and architectures. Even though some studies outlined in Section.~\ref{sub-sec: bc-metaverse} highlight the importance of blockchain in a metaverse, they often remain at the theoretical level without considering specific system implementations. Furthermore, these existing works overlook the CLA and RC issues introduced in Section.~\ref{sec:introduction}, which seriously hinder the development and implementation of a practical metaverse. 
To address these concerns, we propose an adaptive decentralized metaverse based on modular blockchain technologies. Our metaverse makes use of adaptive consensus/ledger changes and resource virtualization, effectively reducing the entry barriers for users and increasing the system security. 

\section{Preliminaries}~\label{sec:Preliminaries}
In this section, we present the necessary background information on modularized blockchain, Non-Fungible Token (NFT), and off-chain payment channel. 

\subsection{Modularized Blockchain}
%
%\textcolor{blue}{Ye-TODO:check content    ---Yihao 0313}
%GPT
Modularized Blockchain is a type of blockchain architecture that aims to improve scalability, flexibility, and interoperability by partitioning different components of a blockchain into modules or layers. In a modularized blockchain, each module is designed to perform a specific function such as consensus, data storage, or smart contract, and can be developed and updated independently~\cite{xu2023exploring}.

A modularized architecture enables a blockchain to adapt to different use cases and business requirements by allowing developers to customize and plug-in different modules. For example, a business can choose a specific consensus algorithm or data storage module that suits its own needs, rather than being restricted to the default options of a blockchain platform.
Moreover, modularized blockchains facilitate interoperability between different blockchain networks by enabling modules to communicate with each other through well-defined interfaces. This allows for the exchanges of data and assets between different blockchains, which can improve the efficiency and convenience of cross-chain transactions.
Several blockchain projects such as XuperChain~\cite{XuperChain} and Celestia~\cite{Celestia} have adopted a modularized blockchain architecture to achieve better scalability and interoperability. 

\subsection{Non-Fungible Token}
Non-Fungible Token (NFT)~\cite{ali2023review} is a digital asset that represents ownership or proof of authenticity of a unique item or a piece of content. Unlike cryptocurrencies such as Bitcoin, which are interchangeable and have the same value, NFTs are unique and each has its own distinct value. NFTs are based on blockchain technologies and are created using smart contracts that verify the ownership and uniqueness of the asset. The concept of NFT has gained significant attention in recent years, particularly in the art world, where NFTs have been used to authenticate and sell digital arts. NFTs provide a way for artists to monetize their digital creations, as the ownership of an NFT can be transferred between buyers and sellers in a secure and transparent way. Additionally, NFTs have the potential to revolutionize the gaming industry by allowing players to truly own in-game items and assets. %Each NFT has a unique identifier and can represent anything from digital art, music, videos, and even virtual real estate in the metaverse.

The process of using an NFT involves creating or purchasing the NFT, which is typically done through a blockchain platform such as Ethereum that supports NFTs. An NFT is created via a smart contract that includes details about the item or content being represented, such as its ownership, provenance, and any additional term or condition of the sale or transfer. 
Once the NFT is created, it can be sold or transferred between individuals, typically through an online marketplace or auction. The transaction is recorded on the blockchain, which ensures that the ownership and authenticity of the NFT are transparent and immutable.
Owners of NFTs can display or store them in digital wallets or galleries, and may be able to earn revenue from their NFTs through licensing or resale. The popularity of NFTs has grown rapidly in recent years, with a few high-profile sales reaching millions of dollars~\cite{ito2022predicting, far2022review}. 

\subsection{Off-chain Payment Channel}
%
%\textcolor{blue}{Yihao-TODO:add content    ---Yihao 0312}
The concept of off-chain payment channels was first proposed by Joseph Poon and Thaddeus Dryja in their paper talking about Lightning Network~\cite{poon2016bitcoin}. It is a technique used in blockchain technologies to enable faster and cheaper transactions between two parties without having to wait for their transactions to be recorded on a blockchain. This is achieved by creating a temporary payment channel between the two parties off-chain. 

\newtheorem{mydef}{Definition}
\begin{mydef}[The Process of Off-chain Payment Channel]
	An off-chain payment channel involves collaborative interactions between on-chain and off-chain. The whole process can be divided into three steps: Open, Transfer, and Close. 
\end{mydef}

\begin{itemize}
		\item  Open (on-chain). Both parties initiate a request to a smart contract separately via transactions to apply for opening a channel and deposit a certain amount of funds. The smart contract verifies the correctness of the transaction signatures and deposit amounts, then sets the deposited amounts from both parties as the initial state of the channel. At this point, the channel is opened.

		\item Transfer (off-chain). Both parties can send transactions back and forth within the channel multiple times. Each transaction needs to be executed by both parties to update the channel state, and the updated state needs to be signed by both parties, which is called a transaction certificate. A transaction certificate includes a timestamp that records the time at which the channel state was updated. 

		\item Close (on-chain). When the interactions between the two parties are complete and the channel needs to be closed, one party (agreed by both parties) can upload the final transaction certificate to the smart contract. The smart contract then verifies the correctness of the signatures and the final state on the certificate, subsequently updating the corresponding state on the blockchain. If the selected party behaves maliciously and fails to upload the correct final transaction certificate (e.g., one provides a certificate that is biased towards its own interest rather than reflecting the final result), the other party can appeal by providing the correct certificate within a designated time frame. The smart contract would verify the authenticity of the certificates by cross-checking the signatures and timestamps provided by both parties. 
\end{itemize}
 
Off-chain payment channels only support off-chain exchanges of currency, which limits their application scenarios. To overcome this restriction, various relevant schemes such as off-chain state channels~\cite{dziembowski2018general}, cross-chain channels~\cite{guo2022cross}, and privacy-preserving off-chain channels~\cite{yu2022zk} gradually emerged to increase the universality, security, and efficiency of off-chain payment channels.

%\clearpage
%-----------------------------------------------

\section{An Adaptive Decentralized Metaverse} \label{sec:Metaverse}
In this section, we first provide an overview on our proposed metaverse system, then detail the protocol, and finally describe a specific work scenario of the metaverse. 

\subsection{Overview}
%
%图
\begin{figure*}[!htb]
\centering
\centerline{\includegraphics[width=0.75\textwidth]{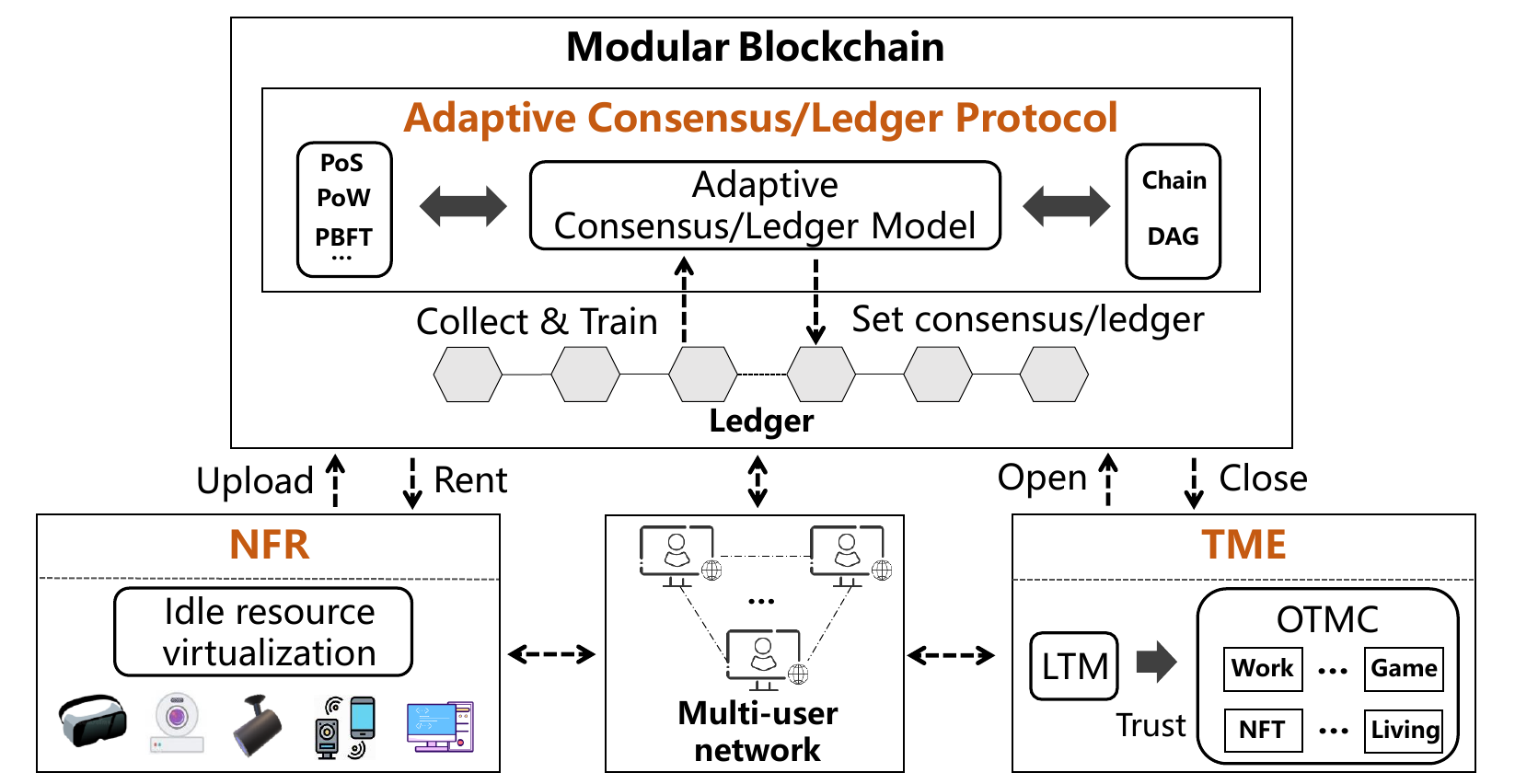}}
\caption{The architecture of a decentralized metaverse. The adaptive consensus/ledger protocols are responsible for collecting data from the metaverse and learning to set up a consensus/ledger that can better suit the scenario. Users can virtualize idle resources as NFR (Non-Fungible Resources) and rent them on the blockchain. The metaverse supports TME (Trusted Metaverse Environment) to realize a locally trusted working environment.} 
\label{fig:overview}
%\vspace{-0.2cm} 
\end{figure*}

As shown in Fig.~\ref{fig:overview}, our proposed metaverse architecture is designed based on a modular blockchain and it mainly consists of three components to support the operations of a metaverse, in which a user can participate through VR glasses or other devices. 

First, considering the challenges that constantly changing metaverse scenes pose on the blockchain consensus and ledger, we propose adaptive consensus/ledger protocols (introduced in Section.~\ref{sec:AMB}) that can automatically determine the most suitable consensus/ledger based on the current state of the entire metaverse, and substitute the consensus/ledger through hot plugging, effectively improving the security of the metaverse and conserving system resources. 
This consideration is based on the modular blockchain design, which allows a low cost consensus change. The hot-pluggable consensus/ledger replacement avoids the deficiency of downtime updates. The introduction of the adaptive protocol with machine learning algorithms reduces the interference of human factors and increases the system's security and robustness, further addressing the CLA problem (as described in Section.~\ref{sec:introduction}). 

Second, a user can virtualize its idle resources, which could be large computers or small sensors, based on our NFR (Non-Fungible Resource) design (introduced in Section.~\ref{sec:NFR}), and upload the NFRs into the metaverse blockchain for rent, maximizing the resource utilization and thereby efficiently addressing the RC problem (as described in Section.~\ref{sec:introduction}). 

%%%%%TME包含LTM+OTMC
Finally, in addition to the adoption of our TME (Trusted Metaverse Environment) design (introduced in Section.~\ref{sec:TME}) to purchase and use NFR, we equip the TME with enhanced security, reliability, and parallel processing abilities through the utilization of an improved Local Trust Model (shown in Section.~\ref{sec:TEOCN}) and an On-Demand Trusted Metaverse Cluster (shown in Section.~\ref{sec:OTMC}). 
Users can join specific metaverse scenes, such as social, NFT, and work, through TME. 
The transaction records of NFRs are written into the blockchain to ensure the correctness of interactions between untrusted parties. 
A large amount of computational tasks are carried out off-chain to reduce on-chain resource consumption. It is worth mentioning that NFR enables users with insufficient resources to join a scene, lowering the threshold for entering the metaverse and thereby fully embodying a people-centric metaverse design. 

\subsection{Adaptive Modular Blockchain} \label{sec:AMB}

To solve the CLA problem, we propose an adaptive consensus protocol and an adaptive ledger protocol based on a modular blockchain. 

\subsubsection{Adaptive Consensus Protocol}
%热插拔共识的含义和意义
Hot-plugging is an ability to replace or install a component without shutting down the computer it attaches. In a metaverse, the complexity of the scenario dictates the need for consensus algorithms to be constantly changing. Therefore, when facing new requirements for performance and security, a developer does not need to design a new blockchain to meet the demands but can upgrade the consensus algorithm online instead, as long as the blockchain system supports hot-pluggable consensus.
 
%热插拔共识机制
A simplified process such as the one in XuperChain to support hot-pluggable consensus can be summarized as follows. First, a user initiates a consensus replacement proposal by calling a smart contract, specifying the name of the consensus algorithm to be replaced and that of the new one, the block height at which the voting would be collected, the trigger condition for the consensus replacement, and the block height at which the new consensus should take effect. Then, other users in the blockchain system vote on the proposal. When the number of votes collected in the smart contract meets the trigger condition for the consensus replacement, the smart contract marks the proposal as ``success'', and all nodes would execute the consensus replacement operation to substitute the consensus, with the new one starting to be effective at the designated block height. If the smart contract fails to collect enough votes, the proposal would be marked as ``failure'', which signals that the consensus update fails and the nodes do not need to perform any operation. 

%自适应共识的意义
However, in a metaverse, users pay more attention to their experience rather than the consensus algorithm adopted by the underlying blockchain system. In fact, the general public may not have any knowledge about consensus. This implies that ordinary users may not be able to initiate consensus proposals following the approach the current hot-pluggable consensus mechanism is implemented (which is more likely a developer's behavior). Therefore consensus changes should be transparent to the users of the metaverse. To realize this objective, we design an adaptive consensus mechanism based on the modular blockchain that supports hot-pluggable consensus and obtain a blockchain that can select an optimal consensus algorithm according to the network scale, number of error nodes, network delay, throughput, and other indicators, to replace the consensus that does not work well due to situation changes, thereby improving the performance and security of the underlying blockchain system. 

%自适应共识
Our adaptive consensus protocol needs a data set that can be obtained by testing the throughput and latency of various consensus algorithms under different network scales, error node ratios, and network delays, over a modular blockchain supporting hot-pluggable consensus. The tags of the data set are various consensus algorithms while the features are indicators such as network scale, error node ratio, and network delay. One can use this data set for training and testing different adaptive consensus models, making use of various machine learning algorithms such as random forest, gradient boosting decision tree (GBDT), XGBoost, and LightGBM~\cite{al2020survey}. The consensus with the highest accuracy is selected and plugged into the blockchain to take effect in the next. 

We summarize the whole process of the adaptive consensus protocol as follows. First, the adaptive consensus protocol obtains the current network state at a fixed interval to identify the optimal consensus algorithm and compares it with the one currently used by the blockchain. If they are the same, the update process stops; otherwise, the protocol broadcasts a transaction (proposal) to the blockchain containing the name of the consensus algorithm to be updated and other relevant information for hot-pluggable consensus change. %the block height at which the new consensus should take effect. 
When a node in the blockchain system receives the transaction, it calls a voting contract, which can count the votes and verify whether the current number of votes has reached the threshold for updating the proposed consensus algorithm. If the threshold is reached, all honest nodes automatically execute the update process of the consensus and broadcast a ``success'' message after the consensus update is successful. If a node fails to complete the update at the specified block height, the node cannot participate in the subsequent consensus and needs to synchronize its status with other nodes. Note that the consensus algorithm library of the modular blockchain is renewable. When someone proposes a new consensus algorithm, the implementation of the algorithm can be added to the library. Therefore, it is necessary to test and count the indicators of the new consensus off-chain, update the data set, and finally retrain the adaptive consensus model.

\subsubsection{Adaptive Ledger Protocol}
%链式账本与图式账本介绍
At present, blockchain ledgers can be categorized into two types: chain-based and DAG-based~\cite{8632193}. In a chain-based ledger, each block, except for the genesis one, stores the hash value of the previous block in its block header. The blocks are then linked together by hashes in a chain-like structure. Examples of prominent blockchains that use a chain-based ledger include Bitcoin~\cite{nakamoto2008bitcoin} and Ethereum~\cite{wood2014ethereum}. In contrast, a DAG-based ledger creates and stores a directed acyclic graph (DAG) of relationships between blocks or transactions. DAG-based ledgers offer higher concurrency than chain-based ones, which results in increased throughput of the blockchain system. However, maintaining consistency in DAG-based ledgers is more challenging, as conflicts take longer time to resolve. Some state-of-the-art DAG-based blockchains include IOTA tangle~\cite{silvano2020iota}, Conflux~\cite{li2020decentralized}, and Sui~\cite{spiegelman2022bullshark}. The choice of ledger implementation should depend on specific application requirements.

\begin{figure}[!htbp]
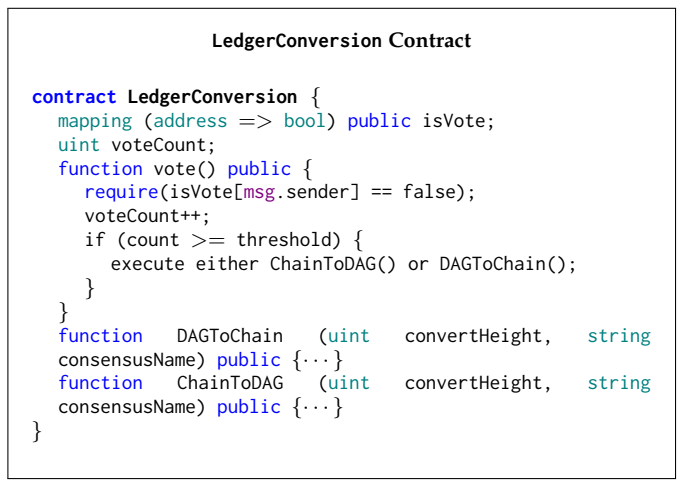
 \footnotesize %\scriptsize
\begin{framed}
\centering \textbf{\texttt{LedgerConversion} Contract}\vspace{10pt}
\texttt{
\begin{basedescript}{\desclabelwidth{20pt}}
    \item[\textcolor{blue}{contract} LedgerConversion] $\{$ \\
    {
        \setlength{\leftskip}{-15pt} \textcolor{teal}{mapping} (\textcolor{teal}{address} $=>$ \textcolor{teal}{bool})  \textcolor{blue}{public} isVote;\\
        \setlength{\leftskip}{-15pt} \textcolor{teal}{uint} voteCount;\\
        \setlength{\leftskip}{-15pt} \textcolor{blue}{function} vote()  \textcolor{blue}{public} $\{$\\
        \setlength{\leftskip}{-5pt} \textcolor{blue}{require}(isVote[\textcolor{violet}{msg}.sender] == false);\\
        \setlength{\leftskip}{-5pt} voteCount++;\\
        \setlength{\leftskip}{-5pt} if (count $>=$ threshold) $\{$\\
        \setlength{\leftskip}{5pt}  execute either ChainToDAG() or DAGToChain();\\
        \setlength{\leftskip}{-5pt} $\}$\\
        \setlength{\leftskip}{-15pt} $\}$\\
        \setlength{\leftskip}{-15pt} \textcolor{blue}{function} DAGToChain (\textcolor{teal}{uint} convertHeight, \textcolor{teal}{string} consensusName) \textcolor{blue}{public} $\{\cdots\}$\\
        \setlength{\leftskip}{-15pt} \textcolor{blue}{function} ChainToDAG (\textcolor{teal}{uint} convertHeight, \textcolor{teal}{string} consensusName) \textcolor{blue}{public} $\{\cdots \}$\\
        \setlength{\leftskip}{-25pt} $\}$\\
    }
    \vspace{2.5pt}   
\end{basedescript}
}
\end{framed}\vspace{-10pt}
\caption{A simplified example of smart contract for ledger conversion.}\label{fig:sc1}
\end{figure}

%提出图式账本与链式账本的转换方法。
We present a novel adaptive ledger protocol that leverages the modular blockchain to enable automated conversion between chain-based and DAG-based ledgers. The conversion process relates to consensus replacement since the two ledger types employ different consensus algorithms. In a manner akin to the adaptive consensus protocol, an additional model is trained for the purpose of facilitating decision-making pertaining to the conversion of ledgers. In order to initiate the conversion process, a miner creates a smart contract as Fig.~\ref{fig:sc1}. The \texttt{LedgerConversion} contract comprises of the following elements: function \texttt{ChainToDAG()}, function \texttt{DAGToChain()}, function \texttt{vote()}, solidity mapping \texttt{isVote}, and uint \texttt{voteCount}. The contract \texttt{LedgerConversion} proceeds in two major steps: voting and ledger conversion. The solidity mapping \texttt{isVote} records the miner address that has cast a vote, thus preventing any miner from voting twice. When a miner invokes \texttt{vote()}, \texttt{voteCount} is incremented after verifying \texttt{isVote}; then the function checks whether the current voteCount has surpassed the threshold. Upon reaching the threshold, nodes all execute \texttt{ChainToDAG()} or \texttt{DAGToChain()}. The call to either function requires the two parameters \texttt{convertHeight} (also denoted by $\hat{H}$ in the following context) and \texttt{consensusName} to be passed, where \texttt{convertHeight} specifies the block height at which the new ledger comes into effect, and \texttt{consensusName} represents the name of the consensus algorithm to be adopted following the conversion.

\begin{figure}[!htbp]
    \centering
    \centerline{\includegraphics[width=0.5\textwidth, height=0.35\textwidth]{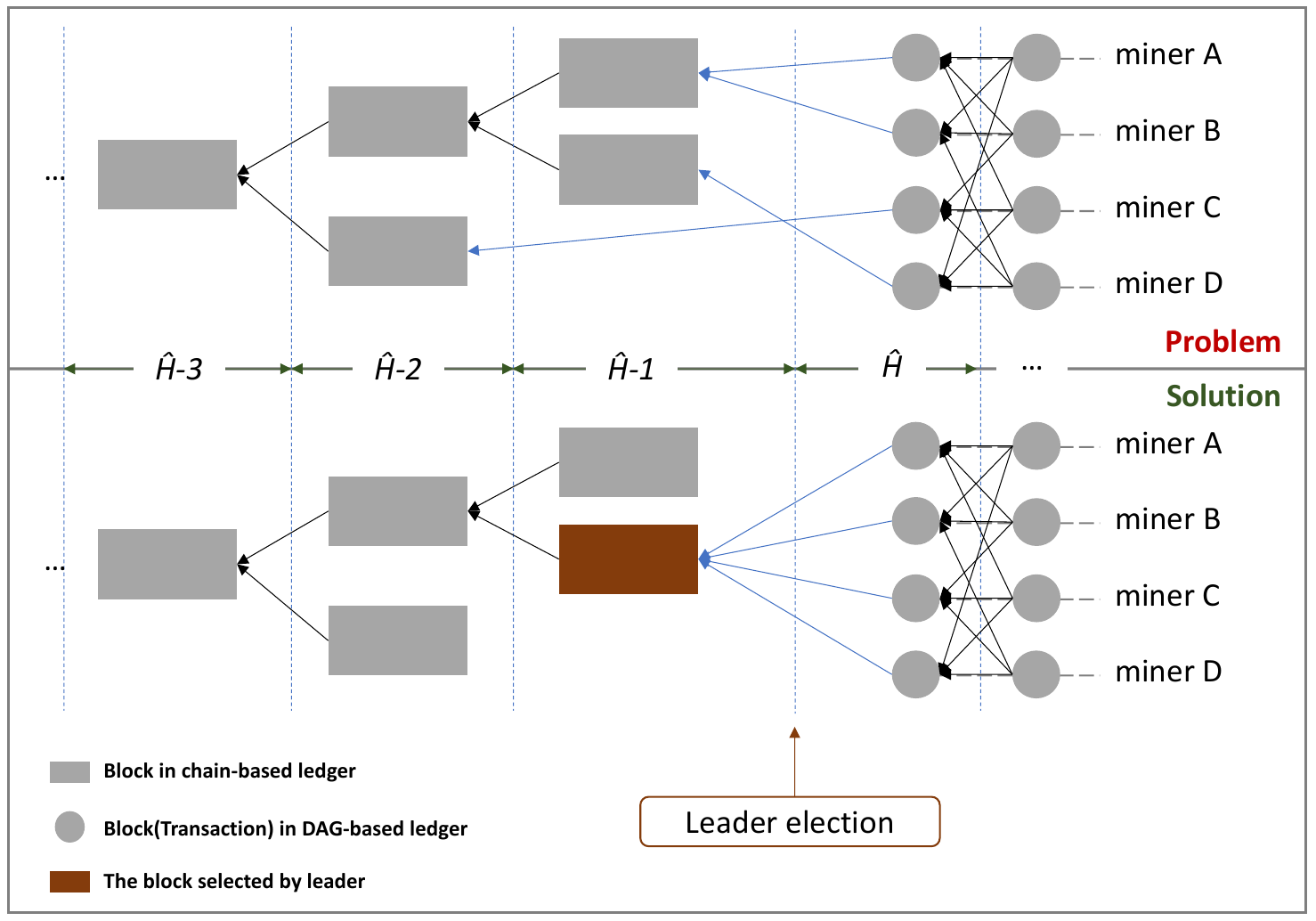}}
	\caption{The challenge and solution concerning the chain-DAG conversion.} 
	\label{fig:ledger1}
\end{figure}

A large number of transactions in a miner's transaction pool can be processed more efficiently by converting a chain-based ledger to a DAG-based structure, thereby increasing concurrency. If a strongly-consistent consensus algorithm such as PBFT is used before ledger conversion, there will be no fork before $\hat{H}$. However, if a weakly-consistent consensus algorithm is used, the block at $\hat{H}-1$ may not be determined when all nodes are in the process of $\hat{H}$. As shown in Fig. \ref{fig:ledger1}, proposed blocks or transactions from different miners at height $\hat{H}$ may point to different branches (forks) before $\hat{H}$, causing inconsistency. To ensure that all miners record the same hash pointer in the proposed blocks or transactions at $\hat{H}$, a distributed randomness beacon is used to conduct leader election between $\hat{H}-1$ and $\hat{H}$. Then the leader selects one path to be committed, and all proposed blocks or transactions at $\hat{H}$ point to the same block, i.e., the leader, ensuring ledger consistency after conversion.

\begin{figure}[!htbp]
    \centering
    \centerline{\includegraphics[width=0.5\textwidth, height=0.25\textwidth]{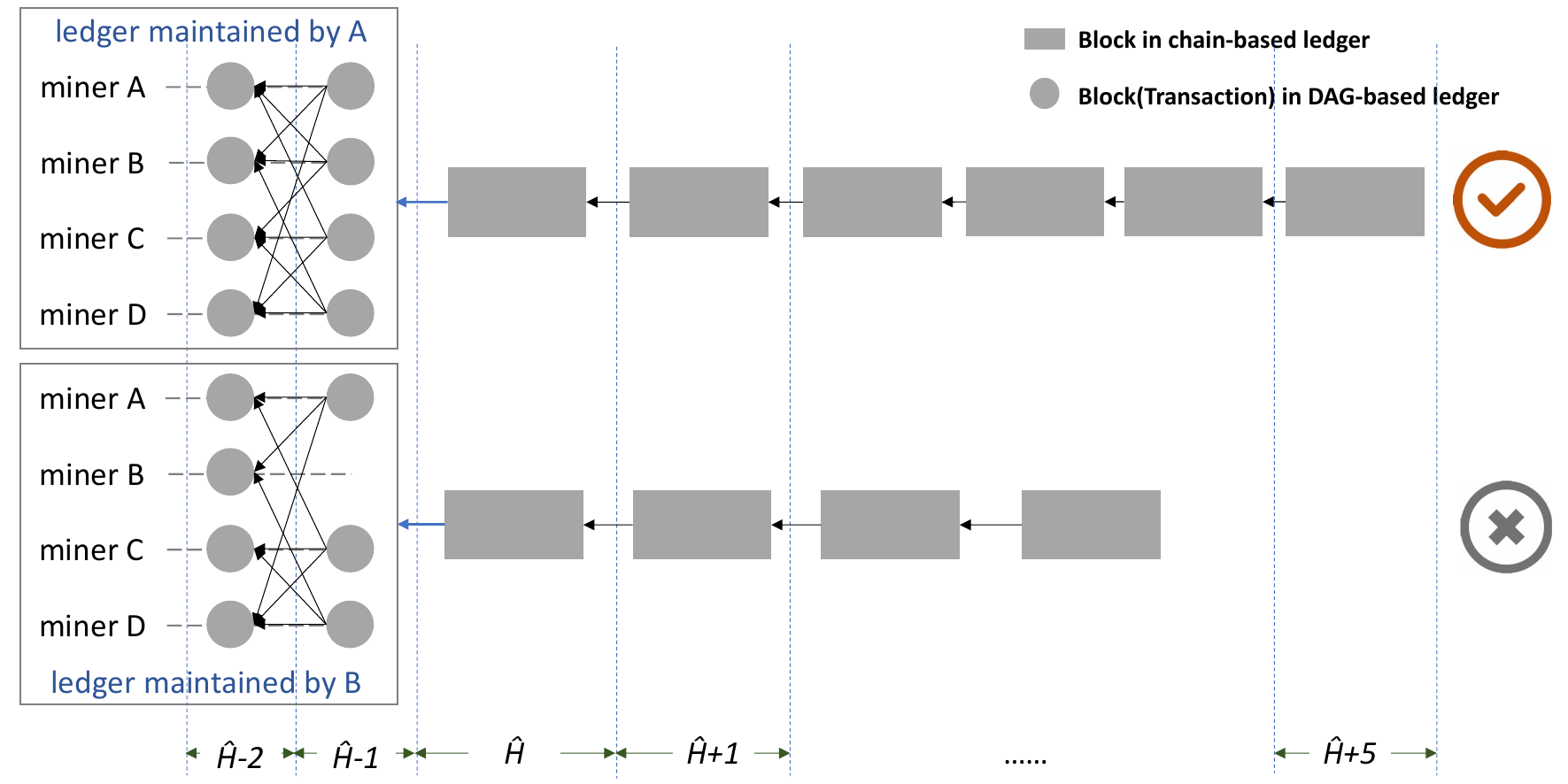}}
	\caption{The challenge and solution concerning the DAG-chain conversion.} 
	\label{fig:ledger2}
\end{figure}

When a miner encounters a few pending transactions in its transaction pool, it can initiate a transaction to invoke DAG-Chain conversion, which can reduce the likelihood of transaction conflicts and strengthen the consistency of the ledger. In the case when the blockchain system operates on a strongly-consistent DAG-based consensus algorithm, there will be no fork at height $\hat{H}$. In the case when the blockchain system utilizes a weakly-consistent DAG-based consensus algorithm, there is a probability that forks might occur in the ledger starting from $\hat{H}$ as depicted in Fig.\ref{fig:ledger2}. Nevertheless, such a fork does not impact the eventual consistency of the ledger after $\hat{H}$ since a chain-based consensus algorithm can ensure the consistency through the longest chain rule or the GHOST rule.

%-----------------------Yihao 1118--------------------------------
%
\subsection{Non-Fungible Resources}
\label{sec:NFR}
A metaverse needs to integrate the computing capabilities of various devices hooked on the Internet. An NFT (Non-Fungible Token) can be understood as a certificate of virtual or physical asset, which is indivisible, irreplaceable, and unique~\cite{nadini2021mapping}. Recently, NFT has been developed rapidly, expanding from games and collectibles to music, real estate, art, and finance~\cite{wang2021non, liu2020tokoin}. 
However, NFT focuses on digital ownership and uniqueness, and it does not support functionalities such as virtualization, access control, and fair-trading of hardware resources, making it incapable of addressing the RC problem. 
To overcome this issue, we propose Non-Fungible Resource (NFR), which can virtualize computational resources including CPU, GPU, and Disk, and map them to tokens recorded by a blockchain.  

We implement NFR using the \texttt{NFR} contract as presented in Fig.~\ref{fig:sc2}. An NFR is described as a solidity struct \texttt{NFR}, including uint \texttt{tokenID}, string \texttt{resourceOwner}, string \texttt{resourceType}, and uint \texttt{price}. The string \texttt{tokenID} is the unique identifier of an NFR. The contract includes the following functions: \texttt{Registration()}, \texttt{Rental()}, \texttt{Liquidation()}, and \texttt{Cancellation()}. The function \texttt{Registration()} is to register the resources from users and returns NFRs to the them. An NFR can only be rented by one user at a time. When a resource-constrained node wants to complete a computing task, it needs to call the function \texttt{Rental()} to rent some NFRs on demand. When calling \texttt{Rental()}, nodes need to provide the rent time and pay a deposit for preventing malicious rentals. If a resource-constrained node completes the computing task, it returns the NFR by calling the function \texttt{Liquidation()}, which automatically settles the cost of rental and return the balance of the deposit to the node. If the node fails to liquidate until the timeout, \texttt{Liquidation()} deduct the deposit from the node and retracts the NFR which is rented by the node. If the owner of a resource no longer rents out its resource forever, it calls the function \texttt{Cancellation()} in the contract to cancel the NFR corresponding to the resource. 
\begin{figure}[!htbp] \footnotesize %\scriptsize
\begin{framed}
\centering \textbf{\texttt{NFR} Contract}\vspace{10pt}
\texttt{
\begin{basedescript}{\desclabelwidth{20pt}}
    \item[\textcolor{blue}{contract} NFR] $\{$ \\
    {
        \setlength{\leftskip}{-15pt} \textcolor{teal}{struct} NFR  $\{$\\
        \setlength{\leftskip}{-5pt} \textcolor{teal}{uint} tokenID;\\
        \setlength{\leftskip}{-5pt} \textcolor{teal}{string} resourceOwner;\\
        \setlength{\leftskip}{-5pt} \textcolor{teal}{string} resourceType;\\
        \setlength{\leftskip}{-5pt} \textcolor{teal}{uint} price;\\
        \setlength{\leftskip}{-15pt} $\}$\\
        \setlength{\leftskip}{-15pt} \textcolor{blue}{function} Registration (\textcolor{teal}{string} resourceType, \textcolor{teal}{uint} price) \textcolor{blue}{public} $\{\cdots\}$\\
        \setlength{\leftskip}{-15pt} \textcolor{blue}{function} Rental (\textcolor{teal}{uint} tokenID, \textcolor{teal}{uint} deposit) \textcolor{blue}{public} $\{\cdots\}$\\
        \setlength{\leftskip}{-15pt} \textcolor{blue}{function} Liquidation (\textcolor{teal}{uint} tokenID) \textcolor{blue}{public} $\{\cdots\}$\\
        \setlength{\leftskip}{-15pt} \textcolor{blue}{function} Cancellation (\textcolor{teal}{uint} tokenID) \textcolor{blue}{public} $\{\cdots \}$\\
        \setlength{\leftskip}{-25pt} $\}$\\
    }
    \vspace{2.5pt}   
\end{basedescript}
}
\end{framed}\vspace{-10pt}
\caption{An example of simplified smart contract of NFR.}\label{fig:sc2}
\end{figure}

Maintaining consistency between digital NFRs and their corresponding physical resources is of utmost importance. Inconsistencies can lead to vulnerabilities, particularly resource fraud attacks. Resource fraud occurs when a user pays for NFRs, but the malicious owners of the NFRs do not provide the corresponding physical resources, resulting in unreliable resources for the user despite payment. To address this issue, we propose a scheme that guarantees the availability and reliability of digital resources corresponding to NFRs in the physical world. We utilize Proofs of Storage (PoS) technique to ensure the validity of storage space when an NFR corresponds to storage resources. PoS allows a user uploading data to a server to repeatedly verify if the server is storing data correctly. For NFRs corresponding to CPUs/GPUs, we employ the Proof of Work (PoW) technique. The server responds to users with a random challenge along with a proof every fixed period of time. If the proof is valid, the resource owner has idle CPU/GPU to serve computing tasks.

In conclusion, our scheme ensures the consistency between digital NFRs and physical resources, thereby mitigating the risk of resource fraud attacks.

\subsection{Trusted Metaverse Environment}~\label{sec:TME}
In this section, we demonstrate the process of enabling parallel computing tasks in a metaverse where users are mutually untrusted. The entire platform consists of two components, with the first assessing the trust of the users involved and the second  constructing an on-demand trusted metaverse cluster.

\subsubsection{Trust Evaluation of Computing Nodes}~\label{sec:TEOCN}

Trust is an important prerequisite for cooperation, which is influenced by historical behaviors and the current state of the nodes. The crisis of trust between nodes has seriously affected the development of distributed systems~\cite{jiang2016understanding, wu2013trust}, and a metaverse is facing similar problems. For example, a node in a metaverse may make a false description of its shared resources to seek improper benefits.

%指明传统的全局现任模型不适合元计算+原因
In a typical metaverse, nodes form different groups according to their computing tasks, and each group works independently. The participation of a large number of heterogeneous devices in the metaverse makes the relationship among all nodes complex and challenging to describe.
% The traditional global trust model (GTM)~\cite{wang2020survey,ahmed2019trust,sherchan2013survey,govindan2011trust} is suited for simple distributed systems such as a blockchain or a P2P network. In this model, all nodes are required to participate in executing the same protocol and share the same level of trust. This makes the metaverse far from efficient. For instance, a group of honest nodes with high trust may be obligated to follow a Byzantine fault-tolerant protocol without exception, resulting in unnecessary high communication overheads. Due to the diversity and complexity of the metaverse, GTM is not an ideal protocol. Treating all nodes equally in GTM would negatively affect system performance and user experience.
%提出本地信任模型+相比传统的优势
To overcome this difficulty, we adopt the Local Trust Model (LTM) mentioned in ~\cite{xu2022trustless}, which satisfies the \textit{locality} property and can effectively solve the above trust evaluation problem. To be specific, %compared with GTM, 
the locality property of LTM implies that the nodes irrelevant to a metaverse task do not need to join in the trust evaluation of the task. In fact, these irrelevant nodes themselves are unwilling to participate in the task to avoid consuming excess resources. 
%To achieve the above point, LTM provides a rigorous mathematical expression for the metaverse network, which can assign a trust value to each group of nodes and describe trust in a fine-grained manner, rather than providing a unified trust value for the entire network. 

\begin{figure}[htp]
\centering
\centerline{\includegraphics[width=0.4\textwidth]{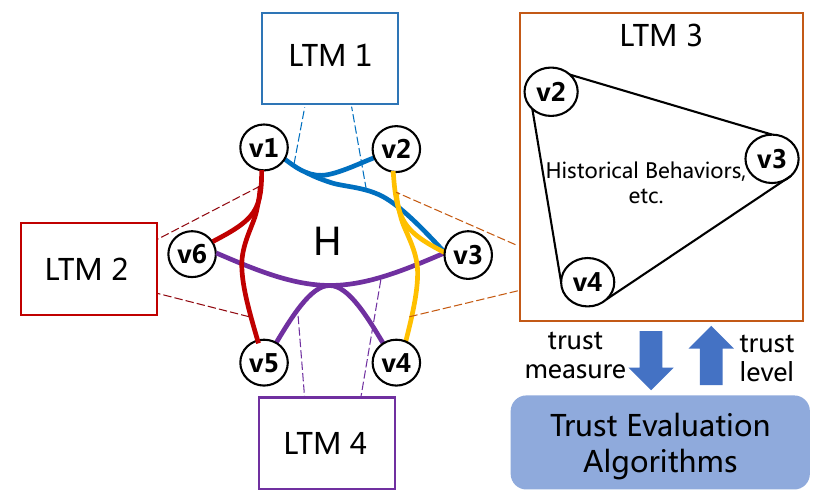}}
\caption{An example of trust evaluation.} 
\label{fig:LTM}
%\vspace{-0.2cm} 
\end{figure}

%本地信任模型解释，超图
Based on the LTM model, we further refine the evaluation criteria and assessment algorithm for the trustworthiness in an LTM. To provide a symbolized expression of the LTM, we denote the metaverse network as a weighted hypergraph~\cite{bretto2013hypergraph} $\mathsf{H\overset{\text{def}}{=}(V, E, W)}$, where $\mathsf{V}$ represents the set of nodes, $\mathsf{E}$ the set of hyperedges, and $\mathsf{W}$ the set of weights. 
Fig.~\ref{fig:LTM} shows an example, which illustrates four LTMs in a metaverse network formed by six nodes (marked as $\{ \mathsf{v1, v2, \cdots, v6 } \}$). One can see that LTM 3 consists of three nodes $\{ \mathsf{v2, v3, v4 } \}$ connected by a hyperedge. 
%考虑数据来源的安全性问题+预言机
The trust value of LTM 3 is determined by the degree of trust among $\{ \mathsf{v2, v3, v4 } \}$, and the factors considered according to specific tasks can include the network size, the message latency, and the historical behaviors of the nodes. We adopt the Oracle mechanism~\cite{Oracle} to aggregate the metaverse network data to calculate trust.
%Furthermore, we consider the problem of malicious nodes providing false information, and poor data quality can lead to inaccurate trust estimation.
%Hence we adopt the Oracle mechanism~\cite{Oracle} to aggregate the metaverse network data to calculate trust. Oracles are entities that have been proven that can help the blockchain connect external systems and guarantee the authenticity of the data uploaded to the blockchain~\cite{pasdar2022connect}. Since the Oracle mechanism does not participate in processes such as consensus, it would not affect the decentralization of the entire system. Moreover, some solutions, e.g., Astraea~\cite{adler2018astraea} and the work in~\cite{nelaturu2020public}, can be adopted to solve the single point of failure in Oracles and further improve the security of the system. 
%具体化信任评估算法
After obtaining the data, nodes in the LTM can choose a trust evaluation algorithm, e.g., Powertrust~\cite{kamvar2003eigentrust}, PET~\cite{liang2005pet}, and those in~\cite{josang2007dirichlet, theodorakopoulos2006trust}, from the trust evaluation module and obtain trust values. Significant elements of the trust evaluation process, such as the participating nodes, the selected trust evaluation algorithm, and the trust values, are recorded on the blockchain, and the smart contract allocates appropriate amount of NFRs to the LTM according to its computing task and the trust values. 

%LTM信任分析
\subsubsection{On-Demand Trusted Metaverse Cluster}~\label{sec:OTMC}

\begin{figure}[htp]
\centering
\centerline{\includegraphics[width=0.48\textwidth]{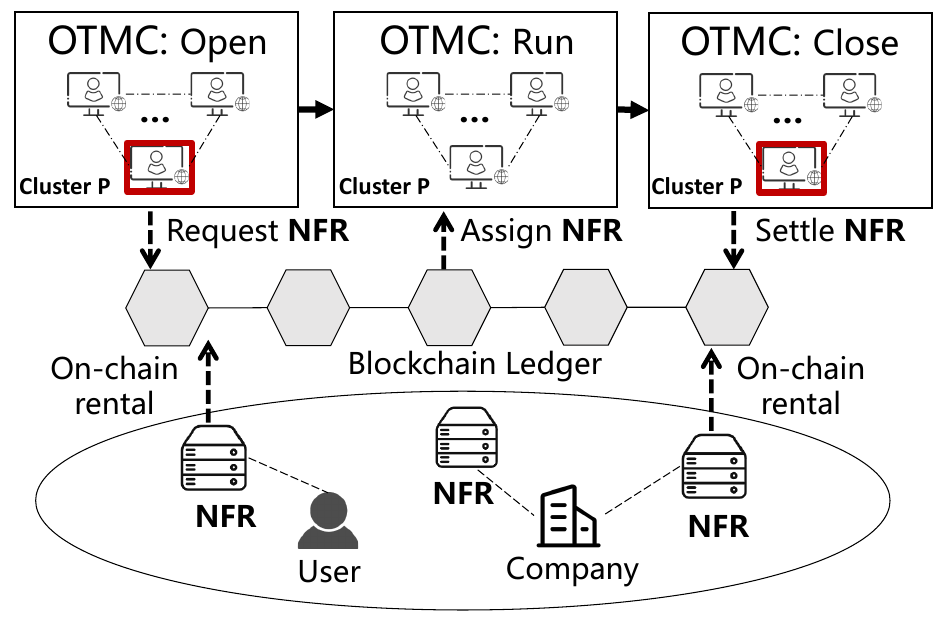}}
\caption{The lifecycle of an OTMC, which contrains three stages: Open, Run, and Close. Nodes with insufficient resources (marked by the red boxes) can participate in a task by renting NFRs.}
\label{fig:OTMC}
%\vspace{-0.2cm} 
\end{figure}

%提出解决方案TMC
Considering that there are many independent and parallel computing tasks in a metaverse, we adopt the On-Demand Trusted Computing Environment technique presented in~\cite{xu2022trustless}. However, due to limitations of the computing power and storage, some nodes are unable to participate. Therefore, we propose the concept of On-Demand Trusted Metaverse Cluster (OTMC) based on NFR (introduced in Section.~\ref{sec:NFR}), which allows users to rent NFRs on the chain to acquire the necessary computing resources, thereby alleviating the resource limitation problem. 
The main idea of OTMC is to create a temporary trusted environment for nodes participating in a task. Note that most of the metaverse tasks are put off-chain to complete, and only some state information of an OTMC needs to be recorded on-chain, such as $\mathsf{Open, Run,}$ and $\mathsf{Close}$ (shown in Fig.~\ref{fig:OTMC}). Multiple computational tasks in the metaverse may occur simultaneously, therefore all OTMCs could be executed in parallel. The participating nodes of each computing task only need to form an OTMC off-chain and rent NFRs on demand to perform the computing task. Since the computing process is off-chain, the process of each OTMC is independent and does not occupy the resources on the blockchain.  

%图

%TMC details
Figure~\ref{fig:OTMC} presents the lifecycle of an OTMC, denoted by $\mathsf{P}$.
We take $\mathsf{P}$ as an example to illustrate the details of an OTMC. 
%TODO：结合NFR
Formally, we denote an OTMC by a vector $\Vec{\mathsf{C}}^\mathsf{State}_\mathsf{Cluster}$$\overset{\text{def}}{=}$$(\mathsf{CID}$, $\mathsf{State}$, $\mathsf{G}$, $\Delta \mathsf{T}$, $\mathsf{NFR}$, $\mathsf{Results})$, where $\mathsf{CID}$ is a unique identifier of the OTMC, $\mathsf{State}$ represents the current state of the OTMC ($\mathsf{State} \in \{\mathsf{Open, Run, Close}\}$), $\mathsf{G}$ represents the participating nodes, $\Delta \mathsf{T}$ is a time duration that regulates how long $\Vec{\mathsf{C}}$ can last, $\mathsf{NFR}$ represents the resource borrowing records of the participating nodes, and $\mathsf{Results}$ is the set of results that needs to be uploaded when $\Vec{\mathsf{C}}$ attempts to be closed. 
Therefore, we can use $\Vec{\mathsf{C}}^\mathsf{Open}_\mathsf{P}$=$(\mathsf{CID_P}, \mathsf{Open}, \mathsf{G_P}, \mathsf{t+\delta}, \mathsf{NFR_P}, \mathsf{\bot})$ to denote the initial state of the cluster $\mathsf{P}$, which states that multiple nodes form a group $\mathsf{G_P}$ and send $\Vec{\mathsf{C}}^\mathsf{Open}_\mathsf{P}$ to the blockchain to establish an on-demand trusted metaverse cluster $\mathsf{P}$ at time $\mathsf{t}$ with a duration $\delta$. Note that, we use the block height to measure $\delta$ to prevent the clock out-of-sync problem among the participants in $\mathsf{G_P}$. 

If $\mathsf{G_P}$ needs to rent resources through NFR, it should deposit certain amount of money in the smart contract. 
Then, it can send requirements, such as computing power consumption and time, to the smart contract. The smart contract would allocate the NFR resources to the corresponding users and record this in $\mathsf{NFR_P}$. 
After the blockchain consensus, the state of $\mathsf{P}$ changes from $\Vec{\mathsf{C}}^\mathsf{Open}_\mathsf{P}$ to $\Vec{\mathsf{C}}^\mathsf{Run}_\mathsf{P}$=$(\mathsf{CID_P}, \mathsf{Run}, \mathsf{G_P}, \mathsf{t+\delta}, \mathsf{NFR_P}, \mathsf{\bot})$, indicating that the participants can start to interact. $\mathsf{P}$ needs to be closed by the time $\mathsf{t+\Delta T}$, otherwise it would be punished (deducting money deposited in the smart contract). Participants upload the results of the interactions to the blockchain, and the miners close $\mathsf{P}$ after successful verification via $\Vec{\mathsf{C}}^\mathsf{Close}_\mathsf{P}$=$(\mathsf{CID_P}, \mathsf{Close}, \mathsf{G_P}, \mathsf{t+\Delta T}, \mathsf{NFR_P}, \mathsf{Results})$. The smart contract settles down the NFR according to the content of $\mathsf{NFR_P}$ and returns the excess amount of the deposit.

\section{IPerformance Evaluation} ~\label{sec:Implementation}
%图
%In this section, we present the concrete implementation of our scheme and test its performance.
In this section, we present a concrete implementation of the major components of our proposed metaverse architecture.

%引用地址
% \noindent{\bf On-chain deployment: Ethereum and smart contract. } The Ethereum Geth\footnote{https://github.com/ethereum/go-ethereum} and Solidity\footnote{https://github.com/ethereum/solidity} come from Github. We use Geth to construct a test network for $\mathsf{Cross}$-$\mathsf{Channel}$ validation, and implement the smart contract $\xi$ based on Solidity.
% In order to facilitate the interactions between smart contract and Ethereum, we adopt web3.py\footnote{https://pypi.org/project/web3} to deploy and call $\xi$.

\subsection{Implementation and Experiment Setup}~\label{Implementation}

%As for the deployment of our scheme, 
We choose XuperChain\footnote{https://github.com/xuperchain/xuperchain} and Solidity\footnote{https://github.com/ethereum/solidity} as building blocks, and implement the adaptive modular blockchain on top of them. Our blockchain system supports three types of consensus algorithms, namely PoW, PoA, and TDPoS. The smart contracts of NFR and OTMC, which are the key components to enable our decentralized metaverse, are realized with Solidity. The adaptive consensus protocol relies on a machine learning model, which intends to choose the most appropriate consensus algorithm based on the number of nodes, the ratio of faulty nodes, and the hardware resources.  %, to build the adaptive consensus protocol. 

% Blockchain network
We test the performance of our adaptive modular blockchain using up to 50 instances distributed in various regions on TencentCloud. In order to compare the impact of different computing power resources on the performance of the consensus algorithms, we set up two networks, which only differ in the amount of computational resources. The first one includes up to 50 S6.MEDIUM4 instances, with each having a 2-Core CPU (Intel Ice Lake 2.7/3.3 GHz), 4GB memory, and 50GB SSD, and running Ubuntu 20.04 LTS; while the second one contains up to 50 C6.LARGE8 instances, with each having a 4-Core CPU (Intel Ice Lake 3.2/3.5 GHz), 8GB memory, and 50GB SSD, and running Ubuntu 20.04 LTS. The bandwidth of each instance is 100 Mbps. Only one XuperChain node is set up on each instance. 

% NFR and OTMC 

 We carry out the following three experiments: 
 \begin{itemize}
 \item Throughput and latency of the modular blockchain.
 \item Latency of the adaptive consensus protocol.
 \item Latency and gas cost of NFR and OTMC operations.
 \end{itemize}
 In the first experiment, we build an XuperChain blockchain system considering three consensus algorithms (PoW, PoA, and TDPoS).  We vary the number of instances added to the blockchain network (from 10 to 50), and send 5000 requests to test the performance of the blockchain system. We record the start and end timestamps of each request and check the blockchain to determine the number of on-chain messages during that period. Using this data, we calculate the TPS (transactions per second) and latency of each request. In the second experiment, we use the results of the first experiment to create a dataset; then we train and test five learning models, i.e., Decision Tree, Random Forest, AdaBoost, XGBoost, and GBDT (all implemented with sklearn and xgboost), and record their accuracy. We select the model with the highest accuracy to build an adaptive consensus protocol. Next we send 20 requests to test the latency of the consensus protocol with varied number of nodes. In the third experiment, we deploy the NFR and OTMC smart contracts on the second blockchain network. Then we designate the nodes on 5 randomly-selected instances as contract invokers to call 8 contract functions 200 times each. %while the nodes on the other remaining instances do not.
 We record the start and end timestamps of each contract function call, as well as the gas consumption, and use the data to calculate the latency and gas of each function in the two contracts.

\subsection{Performance Evaluation Results}~\label{Performance}

\subsubsection{Throughput and Latency of the Modular Blockchain}
This experiment is to test the performance of the blockchain system with respect to different consensus algorithms and to provide training data for the adaptive consensus protocol. In this experiment, we compare three consensus algorithms, i.e., PoW, PoA, and TDPoS, varying the number of nodes, the ratio of faulty nodes, and the hardware resources. The unit for throughput is TPS, which refers to the number of transactions per second. Latency refers to the time delay between initiating a transaction and confirming it on the blockchain.

\begin{figure}[!htbp]
    \subfigure[]{\includegraphics[width=0.24\textwidth]{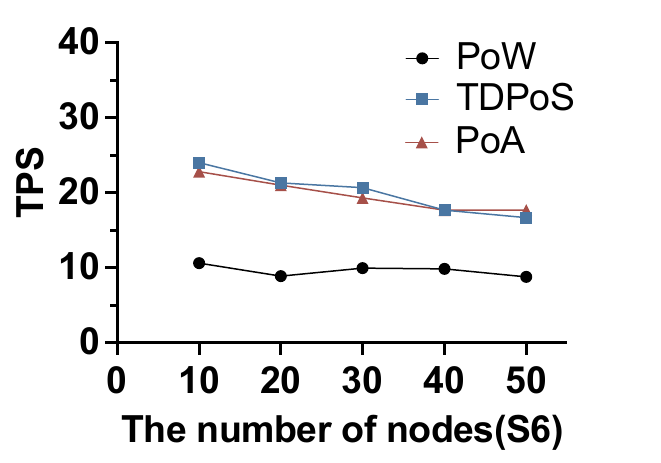}}
	\subfigure[]{\includegraphics[width=0.24\textwidth]{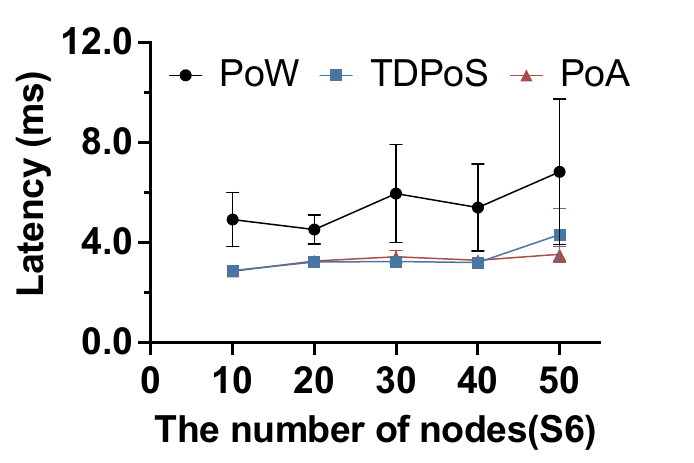}}
	\subfigure[]{\includegraphics[width=0.24\textwidth]{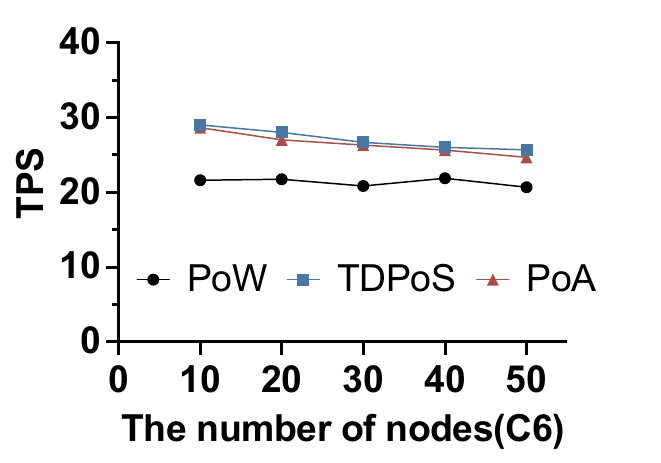}}
	\subfigure[]{\includegraphics[width=0.24\textwidth]{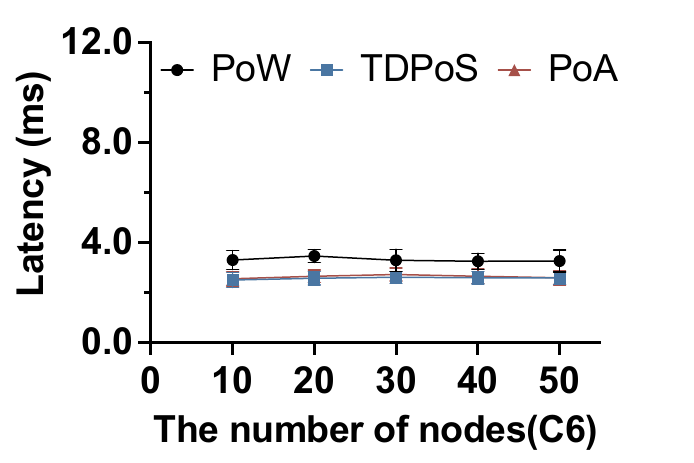}}
	\subfigure[]{\includegraphics[width=0.24\textwidth]{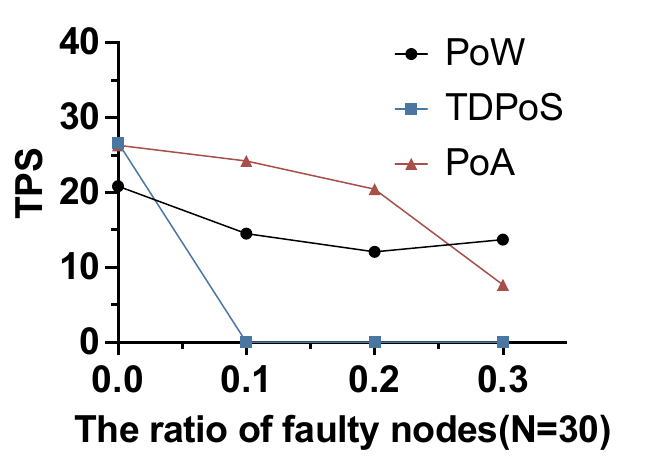}}
    \subfigure[]{\includegraphics[width=0.24\textwidth]{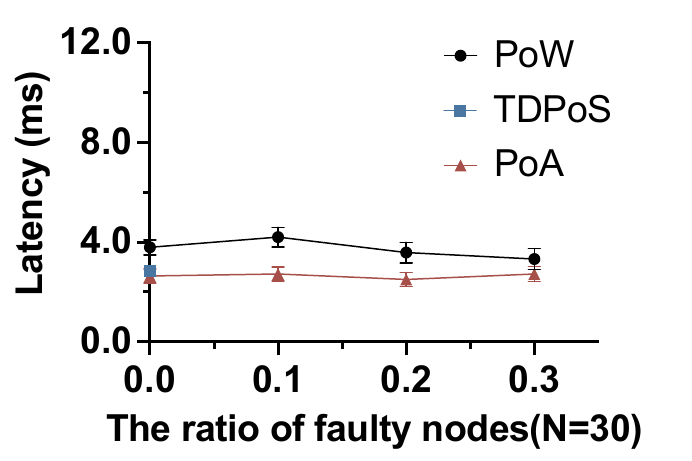}}
	\caption{TPS and latency of different consensus on cloud servers (S6.MEDIUM4) tests (a) (b), TPS and latency of different consensus on cloud server (C6.LARGE8) tests (c) (d),  TPS and latency of different faulty nodes when  the number of nodes is 30 on cloud servers (C6.LARGE8) tests (e) (f).} 
	\label{fig:test1}
\end{figure}

As shown in  Fig.~\ref{fig:test1} (a), the TPS of PoA and TDPoS is better than that of PoW. Specifically, the TPS of the first two consensus algorithms decreases as the number of nodes increases, but that of PoW does not follow this trend. This is because TPDoS and PoA rely on the approvals of sufficient validators to add new blocks to the blockchain, while PoW requires miners to solve a complex mathematical problem. The communication complexity of PoW is relatively low compared to the other consensus algorithms. As the number of nodes increases, the communication time during the consensus process increases, resulting in a TPS decrease of PoA and TDPoS, but the TPS of PoW remains relatively stable. Fig.~\ref{fig:test1} (c) shows that the gap between PoW and the other two consensus algorithms is getting narrower. We find that the node deployed on S6 has less computing power. When the nodes use PoW, the CPU usage tends to reach 100\%, which becomes the bottleneck of performance limitation. When we switch C6 with a more powerful CPU, the bottleneck disappears. 
Fig.~\ref{fig:test1} (b) demonstrates that the latency of PoA and TDPoS remains stable, but the latency of PoW fluctuates wildly. This is also because the weaker CPU limits the performance of PoW. As expected, Fig.~\ref{fig:test1} (d) demonstrates that the latency of PoW remains stable. 

Then we compare the three consensus algorithms with respect to the faulty node ratio ($\mathsf{FR}$), the ratio of the number of existing faulty nodes over the total number of nodes ($N$). The behaviors of the faulty nodes defined here mainly include stopping communications with other nodes, quitting or joining the network at any time, and proposing no blocks when working as a miner. We set $N = 30$, and $\mathsf{FR} \in\{0\%, 10\%, 20\%, 30\%\}$, considering FR $<$ 33\% = f/N, where $\mathsf{FR}$ = 0 represents the non-faulty case. As Fig.~\ref{fig:test1} (e) shows, the TPS of PoA and TDPoS  decreases obviously, and the TPS of TDPoS even decreases to 0 in the presence of $10\% N$  faulty nodes. In Fig.~\ref{fig:test1} (f), as $\mathsf{FR}$ approaches to 10\%, the TPS of TDPoS approaches to 0, and threfore the latency of TDPoS approaches to infinity. When $\mathsf{FR}$ is 30\%, the TPS of PoW surpasses that of PoA. This is because PoA relies on authoritative nodes taking turns to become the leader and propose blocks. When a faulty node becomes the leader, it can cause the blockchain process to stall or slow down. Therefore, as the proportion of faulty nodes increases, the performance of PoA significantly decreases, while that of PoW is not affected. It is clearly shown that the PoW scheme has stronger fault tolerance property.

\begin{table}[!ht]
\centering
\caption{The accuracy of five models}\label{tab:m_acc}
\begin{tabular}{cccccc}
\toprule
\textbf{Models}   & \begin{tabular}[c]{@{}c@{}}Decision \\ Tree\end{tabular} & \begin{tabular}[c]{@{}c@{}}Random \\ Forest\end{tabular} & Adaboost & XGBoost & \textbf{GBDT}   \\
\midrule
\textbf{Accuracy} & 80.0\%        & 68.3\%        & 70.0\%   & 80.0\%  & \textbf{96.7\%} \\
\bottomrule
\end{tabular}
\end{table}

\subsubsection{The Performance of the Adaptive Protocol}
\begin{figure}[!htbp]
    \centering
    \centerline{\includegraphics[width=0.4\textwidth, height=0.17\textwidth]{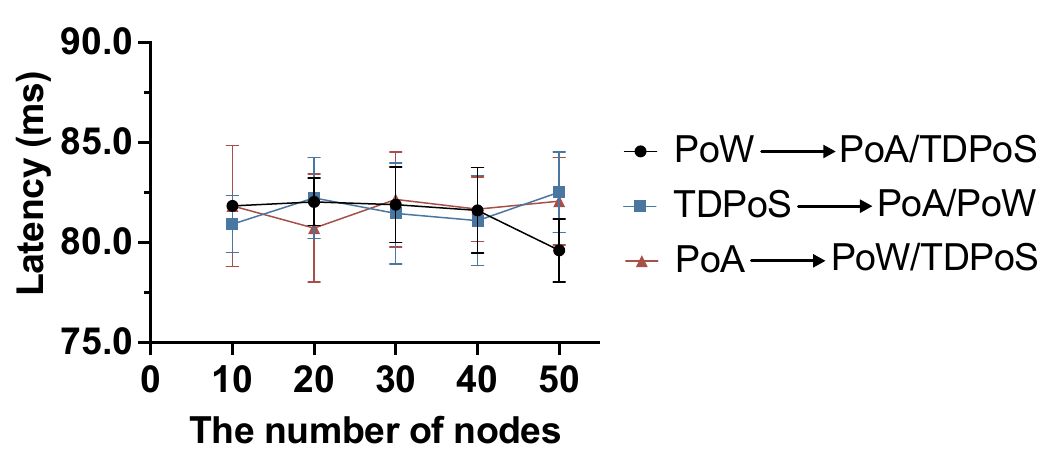}}
	\caption{The latency of updating consensus algorithms.} 
	\label{fig:test2}
\end{figure}

In this experiment, we construct a data set according to the results of the previous experiments. The data set is characterized by the number of nodes and the ratio of the faulty ones. The labels of this dataset include PoA, PoW, and TDPoS. Then, we use this dataset for model training and testing. We use the hold-out method to randomly divide the data set into the training set and the test set according to the ratio of 7:3, and repeat it 10 times, using the average accuracy of all tests as the final evaluation result. We first test the accuracy of Decision Tree, Random Forest, AdaBoost, XGBoost, and GBDT, and the results are shown in Table \ref{tab:m_acc}. One can see that the GBDT model has the highest test accuracy; thus we adopt GBDT as the learning model in our adaptive consensus protocol. Note that the accuracy of these models is measured based on the selected features in the experiment. If more feature information is added to the dataset in future, such as network latency, the model we ultimately select may be different. Then we test the latency of the adaptive consensus protocol. The latency here is defined to be the time duration from when a user inputs parameters into the model, the model outputs results, the input and output are sent as a transaction to the blockchain, until finally the transaction is written into a block. The transaction specifies the type of consensus to be changed and the block height (H) at which the change should occur. Once such information is written into a block, all nodes update their local consensus mechanism at H and notify other nodes of the successful update. As shown in Fig.~\ref{fig:test2}, when the current consensus is PoA, PoW, or TDPoS, the latency of the adaptive replacement to other consensus is similar. This is because, under our experimental conditions, the time to output the model results far exceeds the transaction latency. Therefore regardless of the current consensus, the entire consensus change latency is almost the same.

\subsubsection{Latency and Gas Cost of NFR and OTMC Operations}
\begin{figure}[!htbp]
	\subfigure[]{\includegraphics[width=0.24\textwidth]{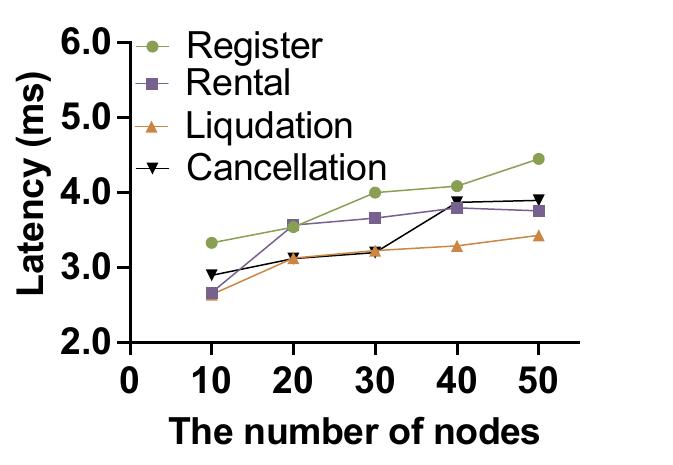}}
	\subfigure[]{\includegraphics[width=0.24\textwidth]{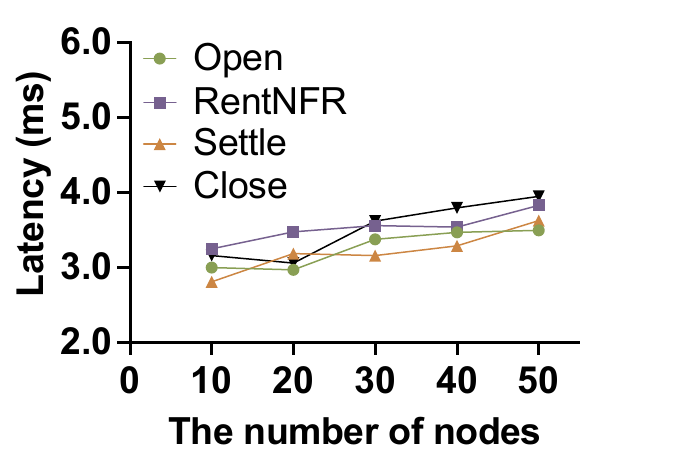}}
	\caption{The latency of NFR and OTMC.} 
	\label{fig:test3}
\end{figure}

\begin{figure}[!htbp]
    \subfigure[]{\includegraphics[width=0.24\textwidth]{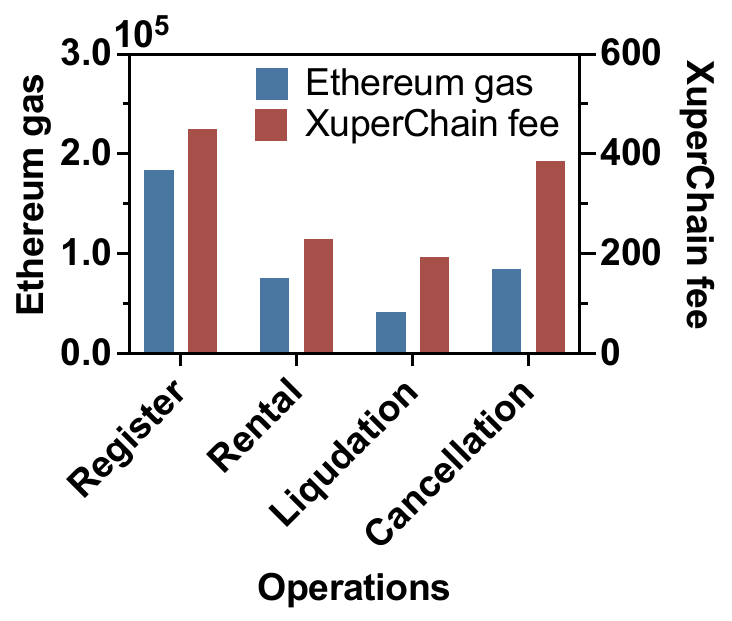}}
	\subfigure[]{\includegraphics[width=0.24\textwidth]{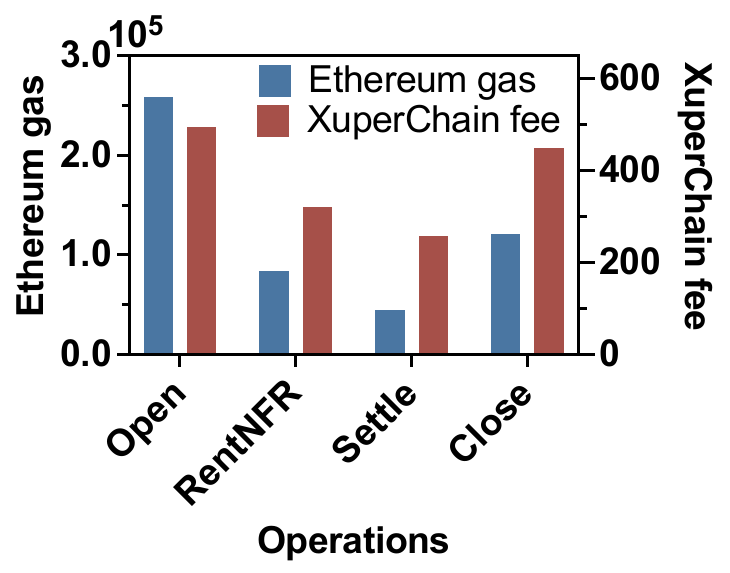}}
	\caption{The gas cost of NFR and OTMC.} 
	\label{fig:test4}
\end{figure}

In this experiment, we test the latency of each function with respect to a variable number of nodes, and the gas consumption of each function in NFR and OTMC. As shown in Fig.~\ref{fig:test3}(a) and (b), one can see that the latency of each function is at the millisecond level, and it increases slightly with the increasing number of nodes. The latency of $\mathsf{Tx_{Open}}$ is the longest, roughly from 3.3ms to 4.5ms, which is reasonable because it involves more uploaded data, e.g., multiple signatures and addresses. We test the Ethereum gas cost and XuperChain fee for each function. As our experiment is based on XuperChain, we also deploy these contracts on Ethereum to facilitate comparisons and provide readers with a better understanding of the specific cost of each contract function. However, as Ethereum and XuperChain utilize different fee calculation methods, the results are not proportional. From Fig.~\ref{fig:test4}(a) and (b), one can see that the gas cost of each function fluctuates between 50,000 and 250,000, and $\mathsf{Tx_{Open}}$ consumes the most gas (about 250,000). The reason lies in that this operation requires more input data and a more complex calculation process. 
% \begin{table}[!htbp]
% 	\centering
%  	\caption{The accuracy of different models experiments}\label{tab:m_acc}
% 	%\begin{threeparttable} \scriptsize
% 		\centering
%   \setlength{\tabcolsep}{10mm}{
% 		\begin{tabular}{cccc}
% 			\toprule
% 			Models  & Accuracy \\
% 			\midrule
% 			$\mathsf{{Decision \ Tree}}$  & 80.0\% \\ % & \fullcirc/ \fullcirc/ \fullcirc   \\
% 			%+4800
% 			$\mathsf{{Random \ Forest}}$ & 68.3\%  \\ % & \emptycirc/ \emptycirc/ \fullcirc  \\
% 			$\mathsf{{Adaboost}}$ & 70.0\%  \\ % & \emptycirc/ \emptycirc/ \fullcirc \\
% 			$\mathsf{{XGBoost}}$ &  80.0\% \\ %  & \fullcirc/ \fullcirc/ \fullcirc \\
% 			$\mathsf{{GBDT}}$ & 96.7\%  \\ % &  \emptycirc/ \emptycirc/ \fullcirc  \\
% 			$\mathsf{{LightGBM}}$ & 35.0\%  \\ % & \fullcirc/ \fullcirc/  \fullcirc \\
% 			\bottomrule
% 		\end{tabular}
%   }
% 	%\end{threeparttable}

% \end{table}

\section{Conclusion} ~\label{sec:conclusion}

In this paper, we present an adaptive and modular blockchain-enabled architecture for a decentralized metaverse. Our architecture tackles two key challenges: the Consensus/Ledger Adaptation (CLA) problem and the Resource Centralization (RC) problem.
To address the CLA problem, we propose an adaptive consensus/ledger protocol that maintains consensus across a distributed network while adapting to changing conditions.
To solve the RC problem, we introduce the concept of Non-Fungible Resource to enable resource virtualization and consolidation within the metaverse without risking centralization.
Additionally, we design the On-Demand Trusted Metaverse Cluster to support parallel computing and promote fair trading of the Non-Fungible Resources.
Finally, we test the performance of our design using three consensus algorithms, and the experimental results demonstrate the feasibility of our approach.

\section{Acknowledgement}
This study was partially supported by the National Key R\&D Program of China (No.2022YFB4501000), the National Natural Science Foundation of China (No.62232010), Shandong Science Fund for Excellent Young Scholars (No.2023HWYQ-008), and the Shandong Science Fund for Key Fundamental Research Project (ZR2022ZD02).

\bibliographystyle{IEEEtran}
\bibliography{references}

% Generated by IEEEtran.bst, version: 1.14 (2015/08/26)
\begin{thebibliography}{10}
\providecommand{\url}[1]{#1}
\csname url@samestyle\endcsname
\providecommand{\newblock}{\relax}
\providecommand{\bibinfo}[2]{#2}
\providecommand{\BIBentrySTDinterwordspacing}{\spaceskip=0pt\relax}
\providecommand{\BIBentryALTinterwordstretchfactor}{4}
\providecommand{\BIBentryALTinterwordspacing}{\spaceskip=\fontdimen2\font plus
\BIBentryALTinterwordstretchfactor\fontdimen3\font minus
  \fontdimen4\font\relax}
\providecommand{\BIBforeignlanguage}[2]{{%
\expandafter\ifx\csname l@#1\endcsname\relax
\typeout{** WARNING: IEEEtran.bst: No hyphenation pattern has been}%
\typeout{** loaded for the language `#1'. Using the pattern for}%
\typeout{** the default language instead.}%
\else
\language=\csname l@#1\endcsname
\fi
#2}}
\providecommand{\BIBdecl}{\relax}
\BIBdecl

\bibitem{ning2021survey}
H.~Ning, H.~Wang, Y.~Lin, W.~Wang, S.~Dhelim, F.~Farha, J.~Ding, and
  M.~Daneshmand, ``A survey on metaverse: the state-of-the-art, technologies,
  applications, and challenges,'' \emph{arXiv preprint arXiv:2111.09673}, 2021.

\bibitem{wang2022survey}
Y.~Wang, Z.~Su, N.~Zhang, R.~Xing, D.~Liu, T.~H. Luan, and X.~Shen, ``A survey
  on metaverse: Fundamentals, security, and privacy,'' \emph{IEEE
  Communications Surveys \& Tutorials}, 2022.

\bibitem{lee2021all}
L.-H. Lee, T.~Braud, P.~Zhou, L.~Wang, D.~Xu, Z.~Lin, A.~Kumar, C.~Bermejo, and
  P.~Hui, ``All one needs to know about metaverse: A complete survey on
  technological singularity, virtual ecosystem, and research agenda,''
  \emph{arXiv preprint arXiv:2110.05352}, 2021.

\bibitem{dionisio20133d}
J.~D.~N. Dionisio, W.~G.~B. III, and R.~Gilbert, ``3d virtual worlds and the
  metaverse: Current status and future possibilities,'' \emph{ACM Computing
  Surveys (CSUR)}, vol.~45, no.~3, pp. 1--38, 2013.

\bibitem{nakamoto2008bitcoin}
S.~Nakamoto, ``Bitcoin: A peer-to-peer electronic cash system,''
  \emph{Decentralized Business Review}, p. 21260, 2008.

\bibitem{fu2022survey}
Y.~Fu, C.~Li, F.~R. Yu, T.~H. Luan, P.~Zhao, and S.~Liu, ``A survey of
  blockchain and intelligent networking for the metaverse,'' \emph{IEEE
  Internet of Things Journal}, 2022.

\bibitem{xu2022trustless}
M.~Xu, Y.~Guo, Q.~Hu, Z.~Xiong, D.~Yu, and X.~Cheng, ``A trustless architecture
  of blockchain-enabled metaverse,'' \emph{High-Confidence Computing}, p.
  100088, 2022.

\bibitem{XuperChain}
\BIBentryALTinterwordspacing
X.~Lab. What is xuperchain. [Online]. Available:
  \url{https://github.com/xuperchain/xuperchain}
\BIBentrySTDinterwordspacing

\bibitem{wood2014ethereum}
G.~Wood \emph{et~al.}, ``Ethereum: A secure decentralised generalised
  transaction ledger,'' \emph{Ethereum project yellow paper}, vol. 151, no.
  2014, pp. 1--32, 2014.

\bibitem{Roblox}
\BIBentryALTinterwordspacing
R.~Corporation. (2006) Roblox. [Online]. Available:
  \url{https://developer.roblox.com/en-us/}
\BIBentrySTDinterwordspacing

\bibitem{Omniverse}
\BIBentryALTinterwordspacing
N.~Team. (2021) Nvidia omniverse. [Online]. Available:
  \url{https://www.nvidia.com/en-us/omniverse/}
\BIBentrySTDinterwordspacing

\bibitem{MicrosoftMesh}
\BIBentryALTinterwordspacing
M.~Team. (2022) Microsoft mesh. [Online]. Available:
  \url{https://www.microsoft.com/en-us/mesh}
\BIBentrySTDinterwordspacing

\bibitem{HorizonWorkroom}
\BIBentryALTinterwordspacing
A.~Heath. (2004) Inside facebook’s metaverse for work. [Online]. Available:
  \url{https://www.theverge.com/2021/8/19/22629942/facebook-workroomshorizon-oculus-vr}
\BIBentrySTDinterwordspacing

\bibitem{secondlife}
\BIBentryALTinterwordspacing
L.~Labs. (2003) Second life. [Online]. Available: \url{https://secondlife.com/}
\BIBentrySTDinterwordspacing

\bibitem{Minecraft}
\BIBentryALTinterwordspacing
M.~Team. (2009) Minecraft maps. [Online]. Available:
  \url{https://www.minecraftmaps.com/tags/real-cities-in-minecraft}
\BIBentrySTDinterwordspacing

\bibitem{Jon2021metaverse}
\BIBentryALTinterwordspacing
J.~Radoff. (2021) The metaverse value-chain. [Online]. Available:
  \url{https://medium.com/building-the-metaverse/the-metaverse-value-chain-afcf9e09e3a7}
\BIBentrySTDinterwordspacing

\bibitem{nair2022going}
V.~Nair, G.~M. Garrido, and D.~Song, ``Going incognito in the metaverse,''
  \emph{arXiv preprint arXiv:2208.05604}, 2022.

\bibitem{duan2021metaverse}
H.~Duan, J.~Li, S.~Fan, Z.~Lin, X.~Wu, and W.~Cai, ``Metaverse for social good:
  A university campus prototype,'' in \emph{Proceedings of the 29th ACM
  International Conference on Multimedia}, 2021, pp. 153--161.

\bibitem{kye2021educational}
B.~Kye, N.~Han, E.~Kim, Y.~Park, and S.~Jo, ``Educational applications of
  metaverse: possibilities and limitations,'' \emph{Journal of Educational
  Evaluation for Health Professions}, vol.~18, 2021.

\bibitem{Decentraland}
\BIBentryALTinterwordspacing
Decentraland. (2017) Welcome to decentraland. [Online]. Available:
  \url{https://decentraland.org/}
\BIBentrySTDinterwordspacing

\bibitem{Cryptovoxels}
\BIBentryALTinterwordspacing
Cryptovoxels. (2021) Welcome to voxels - a user-owned virtual world. [Online].
  Available: \url{https://www.cryptovoxels.com/}
\BIBentrySTDinterwordspacing

\bibitem{yang2022fusing}
Q.~Yang, Y.~Zhao, H.~Huang, Z.~Xiong, J.~Kang, and Z.~Zheng, ``Fusing
  blockchain and ai with metaverse: A survey,'' \emph{IEEE Open Journal of the
  Computer Society}, vol.~3, pp. 122--136, 2022.

\bibitem{huynh2023blockchain}
T.~Huynh-The, T.~R. Gadekallu, W.~Wang, G.~Yenduri, P.~Ranaweera, Q.-V. Pham,
  D.~B. da~Costa, and M.~Liyanage, ``Blockchain for the metaverse: A review,''
  \emph{Future Generation Computer Systems}, 2023.

\bibitem{huang2022fusion}
H.~Huang, X.~Zeng, L.~Zhao, C.~Qiu, H.~Wu, and L.~Fan, ``Fusion of building
  information modeling and blockchain for metaverse: a survey,'' \emph{IEEE
  Open Journal of the Computer Society}, vol.~3, pp. 195--207, 2022.

\bibitem{ersoy2023blockchain}
M.~Ersoy and R.~G{\"u}rfidan, ``Blockchain-based asset storage and service
  mechanism to metaverse universe: Metarepo,'' \emph{Transactions on Emerging
  Telecommunications Technologies}, vol.~34, no.~1, p. e4658, 2023.

\bibitem{ali2023metaverse}
S.~Ali, T.~P.~T. Armand, A.~Athar, A.~Hussain, M.~Ali, M.~Yaseen, M.-I. Joo,
  H.-C. Kim \emph{et~al.}, ``Metaverse in healthcare integrated with
  explainable ai and blockchain: Enabling immersiveness, ensuring trust, and
  providing patient data security,'' \emph{Sensors}, vol.~23, no.~2, p. 565,
  2023.

\bibitem{xu2023exploring}
M.~Xu, Y.~Guo, C.~Liu, Q.~Hu, D.~Yu, Z.~Xiong, D.~Niyato, and X.~Cheng,
  ``Exploring blockchain technology through a modular lens: A survey,''
  \emph{arXiv preprint arXiv:2304.08283}, 2023.

\bibitem{Celestia}
\BIBentryALTinterwordspacing
C.~Lab. The first modular blockchain network. [Online]. Available:
  \url{https://celestia.org/}
\BIBentrySTDinterwordspacing

\bibitem{ali2023review}
O.~Ali, M.~Momin, A.~Shrestha, R.~Das, F.~Alhajj, and Y.~K. Dwivedi, ``A review
  of the key challenges of non-fungible tokens,'' \emph{Technological
  Forecasting and Social Change}, vol. 187, p. 122248, 2023.

\bibitem{ito2022predicting}
K.~Ito, K.~Shibano, and G.~Mogi, ``Predicting the bubble of non-fungible tokens
  (nfts): An empirical investigation,'' \emph{arXiv preprint arXiv:2203.12587},
  2022.

\bibitem{far2022review}
S.~B. Far, S.~M.~H. Bamakan, Q.~Qu, and Q.~Jiang, ``A review of non-fungible
  tokens applications in the real-world and metaverse,'' \emph{Procedia
  Computer Science}, vol. 214, pp. 755--762, 2022.

\bibitem{poon2016bitcoin}
J.~Poon and T.~Dryja, ``The bitcoin lightning network: Scalable off-chain
  instant payments,'' 2016.

\bibitem{dziembowski2018general}
S.~Dziembowski, S.~Faust, and K.~Host{\'a}kov{\'a}, ``General state channel
  networks,'' in \emph{Proceedings of the 2018 ACM SIGSAC Conference on
  Computer and Communications Security}, 2018, pp. 949--966.

\bibitem{guo2022cross}
Y.~Guo, M.~Xu, D.~Yu, Y.~Yu, R.~Ranjan, and X.~Cheng, ``Cross-channel: Scalable
  off-chain channels supporting fair and atomic cross-chain operations,''
  \emph{arXiv preprint arXiv:2212.07265}, 2022.

\bibitem{yu2022zk}
W.~Yu, M.~Xu, D.~Yu, X.~Cheng, Q.~Hu, and Z.~Xiong, ``zk-pcn: A
  privacy-preserving payment channel network using zk-snarks,'' in \emph{2022
  IEEE International Performance, Computing, and Communications Conference
  (IPCCC)}.\hskip 1em plus 0.5em minus 0.4em\relax IEEE, 2022, pp. 57--64.

\bibitem{al2020survey}
M.~A. Al-Garadi, A.~Mohamed, A.~K. Al-Ali, X.~Du, I.~Ali, and M.~Guizani, ``A
  survey of machine and deep learning methods for internet of things (iot)
  security,'' \emph{IEEE Communications Surveys \& Tutorials}, vol.~22, no.~3,
  pp. 1646--1685, 2020.

\bibitem{8632193}
H.~Pervez, M.~Muneeb, M.~U. Irfan, and I.~U. Haq, ``A comparative analysis of
  dag-based blockchain architectures,'' in \emph{2018 12th International
  Conference on Open Source Systems and Technologies (ICOSST)}, 2018, pp.
  27--34.

\bibitem{silvano2020iota}
W.~F. Silvano and R.~Marcelino, ``Iota tangle: A cryptocurrency to communicate
  internet-of-things data,'' \emph{Future Generation Computer Systems}, vol.
  112, pp. 307--319, 2020.

\bibitem{li2020decentralized}
C.~Li, P.~Li, D.~Zhou, Z.~Yang, M.~Wu, G.~Yang, W.~Xu, F.~Long, and A.~C.-C.
  Yao, ``A decentralized blockchain with high throughput and fast
  confirmation,'' in \emph{2020 $\{$USENIX$\}$ Annual Technical Conference
  ($\{$USENIX$\}$ $\{$ATC$\}$ 20)}, 2020, pp. 515--528.

\bibitem{spiegelman2022bullshark}
A.~Spiegelman, N.~Giridharan, A.~Sonnino, and L.~Kokoris-Kogias, ``Bullshark:
  Dag bft protocols made practical,'' in \emph{Proceedings of the 2022 ACM
  SIGSAC Conference on Computer and Communications Security}, 2022, pp.
  2705--2718.

\bibitem{nadini2021mapping}
M.~Nadini, L.~Alessandretti, F.~Di~Giacinto, M.~Martino, L.~M. Aiello, and
  A.~Baronchelli, ``Mapping the nft revolution: market trends, trade networks,
  and visual features,'' \emph{Scientific reports}, vol.~11, no.~1, pp. 1--11,
  2021.

\bibitem{wang2021non}
Q.~Wang, R.~Li, Q.~Wang, and S.~Chen, ``Non-fungible token (nft): Overview,
  evaluation, opportunities and challenges,'' \emph{arXiv preprint
  arXiv:2105.07447}, 2021.

\bibitem{liu2020tokoin}
C.~Liu, M.~Xu, H.~Guo, X.~Cheng, Y.~Xiao, D.~Yu, B.~Gong, A.~Yerukhimovich,
  S.~Wang, and W.~Lv, ``Tokoin: a coin-based accountable access control scheme
  for internet of things,'' \emph{arXiv preprint arXiv:2011.04919}, 2020.

\bibitem{jiang2016understanding}
W.~Jiang, G.~Wang, M.~Z.~A. Bhuiyan, and J.~Wu, ``Understanding graph-based
  trust evaluation in online social networks: Methodologies and challenges,''
  \emph{Acm Computing Surveys (Csur)}, vol.~49, no.~1, pp. 1--35, 2016.

\bibitem{wu2013trust}
X.~Wu, R.~Zhang, B.~Zeng, and S.~Zhou, ``A trust evaluation model for cloud
  computing,'' \emph{Procedia Computer Science}, vol.~17, pp. 1170--1177, 2013.

\bibitem{bretto2013hypergraph}
A.~Bretto, ``Hypergraph theory,'' \emph{An introduction. Mathematical
  Engineering. Cham: Springer}, 2013.

\bibitem{Oracle}
\BIBentryALTinterwordspacing
Chainlink. (2021) What is a blockchain oracle. [Online]. Available:
  \url{https://chain.link/education/blockchain-oracles}
\BIBentrySTDinterwordspacing

\bibitem{kamvar2003eigentrust}
S.~D. Kamvar, M.~T. Schlosser, and H.~Garcia-Molina, ``The eigentrust algorithm
  for reputation management in p2p networks,'' in \emph{Proceedings of the 12th
  international conference on World Wide Web}, 2003, pp. 640--651.

\bibitem{liang2005pet}
Z.~Liang and W.~Shi, ``Pet: A personalized trust model with reputation and risk
  evaluation for p2p resource sharing,'' in \emph{Proceedings of the 38th
  Annual Hawaii International Conference on System Sciences}.\hskip 1em plus
  0.5em minus 0.4em\relax IEEE, 2005, pp. 201b--201b.

\bibitem{josang2007dirichlet}
A.~Josang and J.~Haller, ``Dirichlet reputation systems,'' in \emph{The Second
  International Conference on Availability, Reliability and Security
  (ARES'07)}.\hskip 1em plus 0.5em minus 0.4em\relax IEEE, 2007, pp. 112--119.

\bibitem{theodorakopoulos2006trust}
G.~Theodorakopoulos and J.~S. Baras, ``On trust models and trust evaluation
  metrics for ad hoc networks,'' \emph{IEEE Journal on selected areas in
  Communications}, vol.~24, no.~2, pp. 318--328, 2006.

\end{thebibliography}

\end{document}